# Metasurfaces 3.0: a New Paradigm for Enabling Smart Electromagnetic Environments

Mirko Barbuto, *IEEE Senior Member*, Zahra Hamzavi-Zarghani, Michela Longhi, Alessio Monti, *IEEE Senior Member*, Davide Ramaccia, *IEEE Senior Member*, Stefano Vellucci, *IEEE Member*, Alessandro Toscano, *IEEE Senior Member*, Filiberto Bilotti, *IEEE Fellow*

***Abstract*—So far, the environment has been considered as a source of fading, clutter, blockage, etc., with detrimental consequences for the efficiency and robustness of communication systems. However, the intense research developed toward beyond-5G communications is leading to a paradigm change, in which the environment is exploited as a new degree of freedom and plays an active role in achieving unprecedented system performances. For implementing this challenging paradigm, it has been recently proposed the use of intelligent surfaces able to control almost at will the propagation of electromagnetic waves. In this framework, metasurfaces have emerged as a promising solution, thanks to their field manipulation capabilities achieved through low-cost, lightweight, and planar structures.**

**The aim of this paper is to review some recent applications of metasurfaces and cast them in the scenario of next-generation wireless systems. In particular, we show their potentialities in overcoming some detrimental effects presented by the environment in wireless communications, and discuss their crucial role towards the practical implementation of a smart electromagnetic environment.**

## I. Introduction

IN WIRELESS communications systems, signals captured by the receiver are typically highly affected by the environment (e.g., foliage, buildings, walls, human bodies, etc.), which interacts with the electromagnetic waves through different physical mechanisms, such as reflection, refraction, diffraction, and absorption, and causes degradation to the signal level compared to an ideal Line-of-Sight (LoS) scenario. This limiting effect is expected to grow in 5G and beyond-5G communications, where the mm-wave frequency range will be ever more used [1]. Indeed, the abundant spectrum available in this frequency range allows the support of wide-bandwidth services, such as video streaming, augmented/virtual reality (AR/VR), and vehicular communications. However, compared to sub-6 GHz communications, the system performance are more susceptible to many environmental factors [2]-[4]. Furthermore, the concept of seamless connectivity for everybody and everything, which drives the development of beyond-5G communications, has introduced several new challenges, such as massive capacity, almost-zero latency and jitter, ultra-high reliability, and security. Facing these challenges by using the current 5G technology would require a huge number of access points/base stations and massive local computation. This is especially true in complex indoor environments where thousands of machines, wearable devices, fixed terminal, etc., dynamically exchange information at ultra-high data rate. Solutions based on current technologies may rely on massive MIMO systems, Software Designed Networks (SDN) or Network Virtualized Functions (NVF). However, all of them involve a dramatic increase of computational resources and elaboration time, because of the virtualization of hardware functions at software level. For solving these issues and satisfy the key performance indicators (KPIs) of future wireless systems [5] a paradigm change is, thus, required.

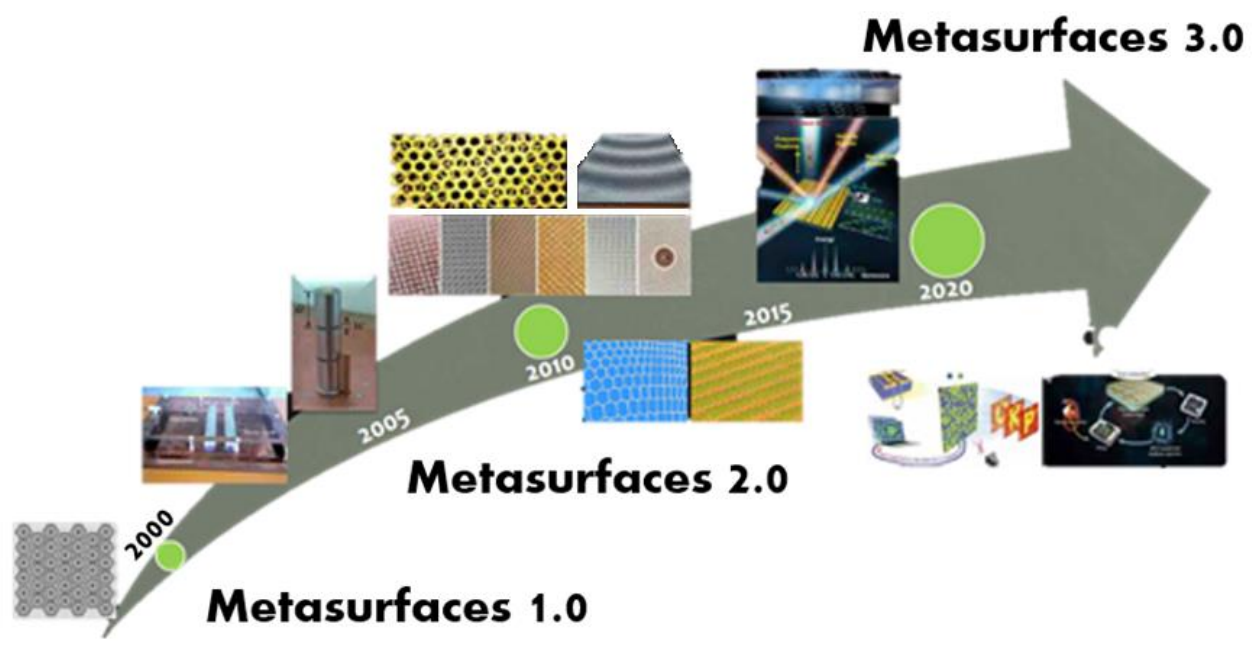

Fig. 1. Time-line of the metasurface technology development: every ten years a new generation of metasurfaces has been introduced with novel and disruptive performances.

In this framework, a significant role as a key enabling technology (KET) for future communication systems can be played by *metasurfaces* [6]. They consist of a planar array of subwavelength-spaced and electrically small particles, whose electromagnetic properties can be locally controlled to tailor the

Manuscript received August XX, 2021This work has been developed in the frame of the activities of the Project MANTLES, funded by the Italian Ministry of University and Research under the PRIN 2017 Program (protocol number 2017BHFZKH). (Corresponding author: Filiberto Bilotti.).

Zahra Hamzavi-Zarghani, Davide Ramaccia, Stefano Vellucci, Alessandro Toscano, and Filiberto Bilotti are with the Department of Industrial, Electronic, and Mechanical Engineering, ROMA TRE University, 00154 Rome, Italy.

Mirko Barbuto, Michela Longhi and Alessio Monti are with the Department of Engineering, Niccolò Cusano University, 00166, Rome, Italy.

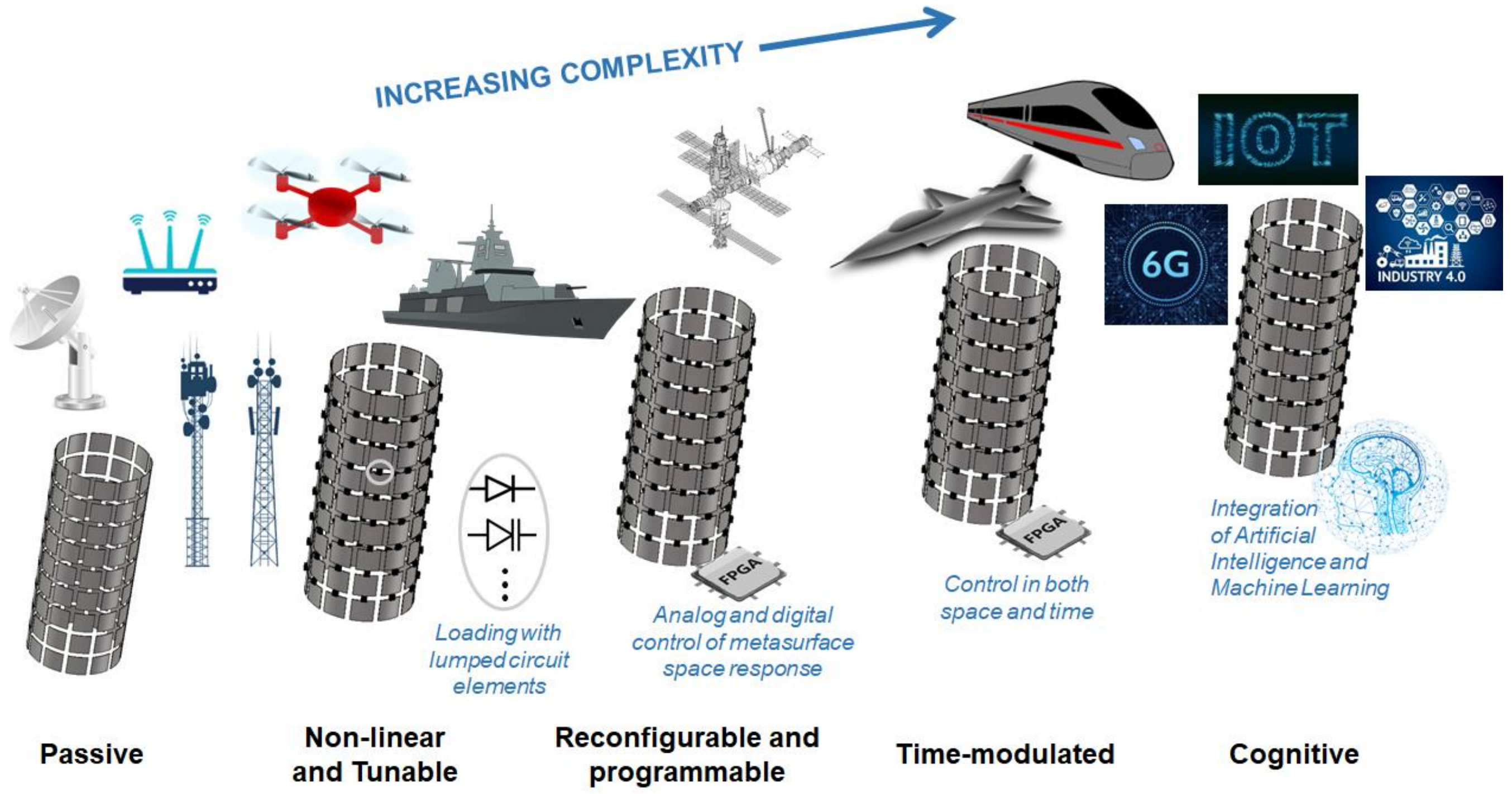

Fig. 2. Schematic representation of the increasing complexity of metasurfaces for different applicative scenarios. By increasing dimensions and movement speed of the objects, as well as, complexity of the operative environment, metasurfaces should be provided by higher degree of freedom, reconfigurable features and cognitive capabilities.

overall reflected/transmitted field. During the last twenty years, many groups worldwide have investigated the possibilities offered by metasurfaces, starting from improving the performances of existing EM components and devices [6]-[18] to inventing conceptually new ones (e.g. cloaks and illusion devices, etc.) [19]-[24], up to enabling tunability, reconfigurability, smartness, and awareness [25]-[36]. The time-line of the metasurface technology is reported in Fig. 1, where we can identify three different generations: the first generation of metasurfaces, here denoted as *Metasurfaces 1.0*, was characterized by regular homogeneous periodic structures, with typical applications in antenna systems (e.g. cloaking devices [19]-[23], lenses [10]-[11], polarization transformers [12]-[13], wide-angle impedance matching sheets [14], spatial filters [15], high-impedance surfaces [16], and artificial magnetic conductors [17], etc.). The second generation of metasurfaces, named *Metasurfaces 2.0*, was instead characterized by inhomogeneous structures with a quasiperiodic or gradient distribution of the amplitude and phase profile across its surface, exhibiting a surface impedance that varies point by point, i.e. spatially modulated metasurfaces [19]. In this case, the surface profile can be engineered for manipulating the reflected and transmitted electromagnetic field, controlling the scattered radiation. These devices enable further applications in antenna systems, such as the design of metasurface antennas [8] or metasurface lenses for pointing the main beam towards anomalous directions [10]. Finally, only very recently, a new degree of freedom has been used for widening the possibility offered by metasurfaces to control the interacting electromagnetic field: the temporal modulation. This led to a third-generation, i.e. *Metasurfaces 3.0*, which refers to either homogenous or inhomogeneous structures whose properties can be controlled in time. The possibility to act on the metasurface properties in both time and space allows increasing the possibilities in designing unique and novel electromagnetic devices, exhibiting tunable, time-modulated, reconfigurable, or programmable responses [25]-[36]. Moreover, the recent paradigm of "Metamaterial-by-Design" has further extended the application of metamaterial and metasurfaces to complex scenarios, by proposing a task-oriented design approach [38]-[40]. Metasurfaces can be, thus, considered an enabling technology for controlling the wireless electromagnetic environment, making it at the service of the communication system. In this framework, the communication environment exhibits the required smartness for supporting the performances to be reached by the 5G and beyond-5G communication system.

The aim of this paper is therefore to present and discuss the potentialities of metasurfaces in overcoming the most deteriorating effects presented by a wireless electromagnetic environment, such as the mutual blockage effect in overcrowded antenna systems, no-LOS communication due to large objects, and time-varying fading due to fast moving scatterers, just to name a few. Moreover, we show that the degrees of freedom offered by metasurfaces make the challenges of beyond-5G communication systems be moved to the physical layer, with the objectives of minimizing hardware virtualization and making the environment itself contributing actively to the system, i.e. conceiving a smart electromagnetic (EM) environment.

The structure of the paper is as follows. According to the schematic representation reported in Fig. 2, which shows the increasing complexity of metasurfaces for different applicative scenarios, in Section II, we show how passive metasurfaces enable innovative scattering-control approaches for radiators and structural elements in complex EM environments. In Section III, we discuss how it is possible overcoming the blockage effects of large objects in LOS microwave links. Then, in Sections IV and V, we show how increasingly adding complexity to the metasurfaces (i.e., by using non-linear, tunable, reconfigurable, or time-modulated structures) it is possible to obtain ever more advanced effects, such as zero fading from moving objects or radiating structures with an environmental-dependent behaviour. Finally, in Section VI, we discuss the possibility to exploit metasurfaces as extended environment coatings able to generate anomalous reflection and refraction to further extend the capabilities of a smart EM environment.

## II. Smart EM environment with Reduced scattering from radiators

The never-ending demand for network resources enforces the use of ever higher frequency bands, massive densification of the site grid, and an increasing co-siting of different antennas. In parallel, the complexity of the radiating elements is progressively growing and multiple-antenna technologies, already adopted in the context of 5G, are expected to play a a key role in the next generation of wireless systems [5].

The overcrowding of the telecommunication platforms introduces several undesired effects occurring at the physical level. Indeed, like any other non-transparent object, also the antenna re-radiates part of the EM energy impinging on it. This phenomenon, known as EM scattering, yields multiple effects, depending on the scenario under consideration (Fig. 3). For instance, when different antennas are co-sited in a limited area, antenna scattering is responsible for both blockage and coupling effects. The former effect results in the deterioration of the radiation patterns of the antennas, while the latter causes a degradation of the quality of the received signals. Similarly, when multiple antenna technologies are considered, antenna scattering causes crosstalk interferences between the array elements that may limit the maximum achievable performances by the system [38].

It is worth noticing that the issues introduced by the undesired antenna scattering are more challenging than the ones caused by the undesired scattering by objects without EM functionalities for several reasons: first, any radiator generally exhibits a higher scattering compared to the one of a passive object with comparable size and composed by the same materials because of the additional contribution caused by the antenna mode [42]. Furthermore, most of the conventional techniques used to reduce the scattering from passive objects, e.g., shaping, radar-absorbing materials and passive/active cancellations, cannot be successfully applied to antennas without affecting their performances. Finally, when an antenna is used in the receiving mode, fundamental bounds apply to the balance between absorption and scattering [43]. Therefore, it is impossible designing a passive antenna able to extract some power from the impinging field whilst exhibiting zero scattering.

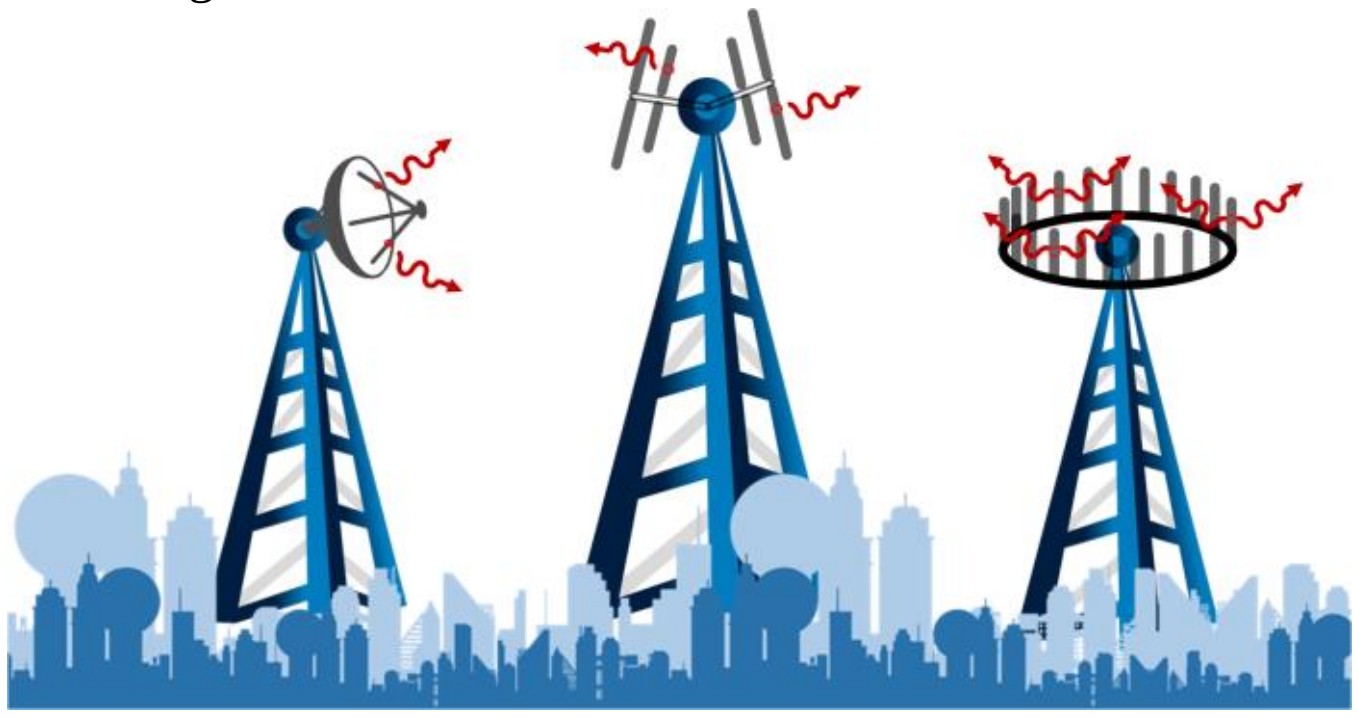

Fig. 3. Illustrative sketch of some of the issues caused by antenna scattering in overcrowded communication platforms: i) blockage from structural elements of the platform; ii) mutual blockage and interference between different radiators; iii) self-interference between radiators in multiple-antenna configurations.

In the above context, the research about antennas exhibiting reduced scattering and limited blockage towards nearby radiators play an important role in the development of the next-generation radio platforms. In the last years, there have been several efforts towards this goal. For instance, it has been shown that a proper combination of electric and magnetic sources allows designing a radiating element exhibiting a Huygens-like pattern with zero back-scattering [44]. Similarly, the use of engineered ground planes has shown the potential to reduce the forward radar cross-section of printed antennas by re-distributing the scattered power in different directions [45]. Unfortunately, none of these techniques ensure a reduction of the total scattering cross-section of the antenna, which is defined as the integral of its radar cross-section over the entire solid angle [42]. As such, these antennas would still introduce blockage, coupling, and self-interference if a specific mutual orientation between nearby radiators could not be ensured.

A radically different possibility for the design of low-scattering antennas suitable for smart EM environments is based on the use of the so-called cloaking techniques [46]. A cloak is defined as a device capable of reducing the total scattering cross-section of an object, without increasing its absorption. This effect can be achieved relying on different techniques but the most suitable one for practical applications involving antennas is the so-called scattering cancellation. In this approach, the invisibility effect is obtained by covering an object with a coat specifically designed to re-radiate a field out-of-phase with the field scattered by the object itself. To achieve such unusual behaviour, the cloaking device should exhibit negative polarizability compared to the one of the object to hide and, thus, can be realized only by relying on either volumetric metamaterials or ultra-thin metasurfaces [47],[19]. The latter solution is preferable given its reduced weight, space occupancy, and easiness of realization [48].

In the following sub-sections, we briefly discuss the innovative degrees of freedom enabled by cloaking at the hardware level and emphasize their potential impact on the design of smart EM environments.

### A. Cloaking of structural elements

As a first possible application of cloaking in a smart EM

environment, we consider the use of metasurfaces to minimize the blockage effects introduced by the structural elements of the communication platforms. As a case study, we discuss the design of a miniaturized satellite (1 cubic unit) equipped with a turnstile crossed-dipole antenna operating in the UHF amateur satellite band (435–438 MHz) and Langmuir probes used to measure the ionosphere electron density [49]-[51]. The probes are installed at the end of four metallic booms placed on the opposite sides of the satellite and deployed after its launch. As it can be appreciated in Fig. 4(a), these structural elements represent a near-field obstacle for the dipolar antenna and degrade significantly its radiation diagram that is expected to be as much isotropic as possible to minimize the average free-space path loss between the space platform and the ground station. However, by covering the metallic booms with a properly engineered metasurface, consisting of a regular lattice of vertical metallic strips placed onto a dielectric layer, it is possible making their scattering cross-section equal to zero within the antenna bandwidth and restoring the original radiation diagram (blue line in Fig. 4(b)). The use of cloaking metasurfaces to minimize the blockage effects of structural elements and obstacles has been also successfully applied in different scenarios involving horn, reflector, and log-periodic antennas [52]-[55]

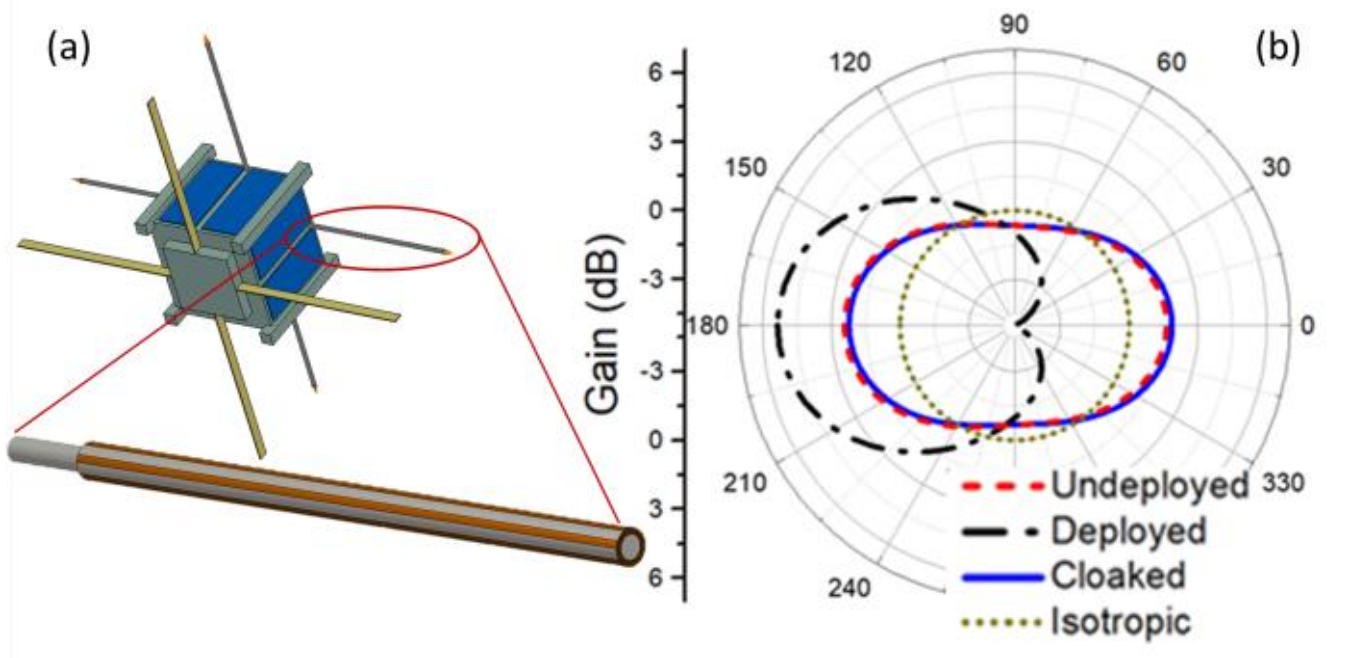

Fig. 4. (a) Miniaturized satellite (1 cubic unit) equipped with a crossed-dipole antenna and Langmuir probes. The metallic booms that support the probes are covered by an inductive metasurface (see the inset) placed onto a dielectric material with low water absorption and low outgassing suitable for space applications; (b) Radiation diagram of the crossed-dipole antenna when the metallic booms are undeployed (red line), deployed and uncloaked (black line), deployed and cloaked (blue line). For comparison, an ideal isotropic radiation diagram is also reported.

## B. Cloaking of co-located antennas

An even more interesting possibility enabled by cloaking is related to the co-siting of different antennas within a limited space. Indeed, by properly engineering the cloaking metasurface, it is possible making an antenna completely transparent at a desired frequency outside its operation bandwidth, whilst maintaining the required radiative and electrical characteristics around its resonance frequency [56],[57]. Such a possibility allows tightly packing different antennas together in a highly dense platform, overcoming the interferences and blockage effects that would occur if cloaking were not used.

As a relevant example, Fig. 5(a) reports a picture of an innovative radiative platform consisting of two monopole antennas working in the frequency range of the third and fourth generation of mobile communications (1900–2200 MHz and 790-860 MHz, respectively). The 4G radiator is made invisible to the 3G antenna within its bandwidth through a metasurface made by horizontal metallic rings placed onto a dielectric substrate. The metasurface is designed to match the cloaking condition [48] within the 3G frequency range while being almost transparent to EM waves within the 4G bandwidth. Consequently, the 4G radiator still works correctly as the metasurface were not present but its blockage effect towards the nearby radiator is almost totally suppressed. In Fig. 5(b), we report the measured radiation diagram of the 3G antenna at $f_0$ = 2100 MHz when it is alone, in presence of the bare 4G radiator, and, finally, when the cloaking metasurface is applied to the 4G antenna. As it can be appreciated, the cloaking metasurface allows the 3G antenna operating as if the electrically long near-field obstacle were not present.

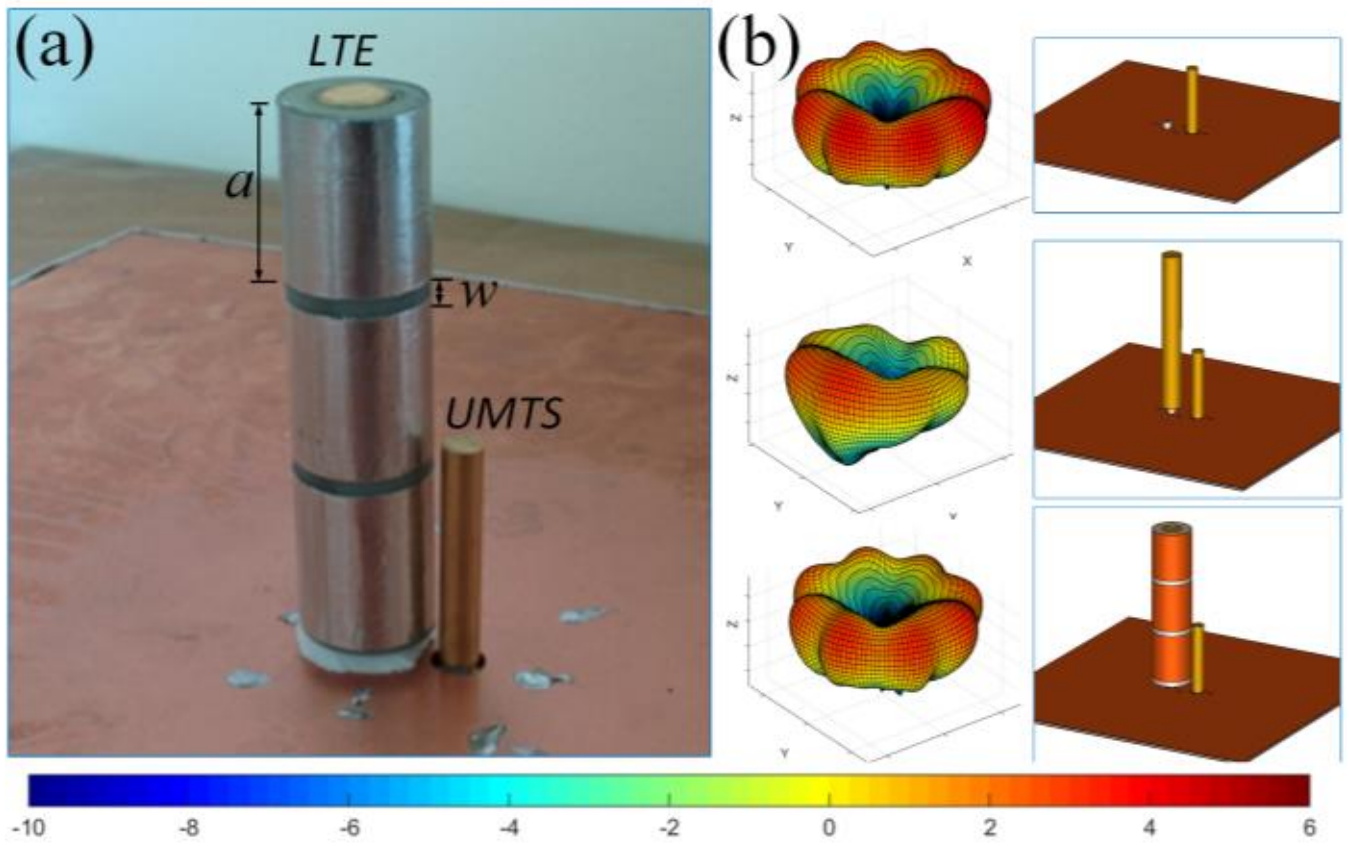

Fig. 5. (a) New-generation radio-platform composed by two tightly-packed monopole antennas working at different frequencies. The 4G antenna is covered by a conformal cloaking metasurface; (b) Measured gain diagram of the 3G monopole in the following scenarios: bare 3G monopole (top), co-located monopoles without cloaking metasurface (center), co-located monopoles with cloaking metasurfaces (bottom).

The possibility to overcome the blockage effects between close antennas has been extended to other radiators and different conditions in [58]-[61]. In general, this approach returns dramatic advantages in terms of minimization of space occupancy and re-use of available radio platforms.

## C. Design of low-scattering antennas

Another innovative degree of freedom enabled by passive cloaking for the next-generation wireless systems relies on the possibility to design antennas exhibiting a low in-band scattering cross-section. For this purpose, it is possible covering an antenna with a metasurface designed to hit the cloaking condition at its own resonance frequency. However, fundamental bounds between absorption and scattering of any passive scatterer apply [62] and, consequently, the reduction of the antenna scattering also implies a reduction of its capability to extract power from the incident field. The advantage of cloaking in this context is the possibility to design antennas exhibiting the best trade-off between absorption and scattering for any desired level of received power [63].

To better explain this point, in Fig. 6 we show the performance of a half-wavelength dipole cloaked with a metasurface able to massively reduce its total scattering at its own resonance frequency. The plot reports the absorption efficiency of the antenna, defined as the ratio between its absorption and scattering cross-sections, as the level of the absorbed power increases. The grey area of the plot is forbidden for passive scatterers [62] and, thus, the boundary between the white and grey areas represents the best operative condition for a receiving antenna. The two lines in the plot are obtained by sweeping the value of the antenna load resistance. As it can be appreciated, in the cloaked scenario the dipole exhibits significantly higher values of the absorption efficiency compared to the bare dipole for the same load resistance. The behaviour of the antenna, thus, can be easily tuned within a wide range of possibilities, such as super-scatterer mode (when a low resistive load is used), maximum absorption mode (when the load resistance equals the radiation resistance), and low-scattering mode (above this resistance value), depending on the desired operating conditions of the individual radiator.

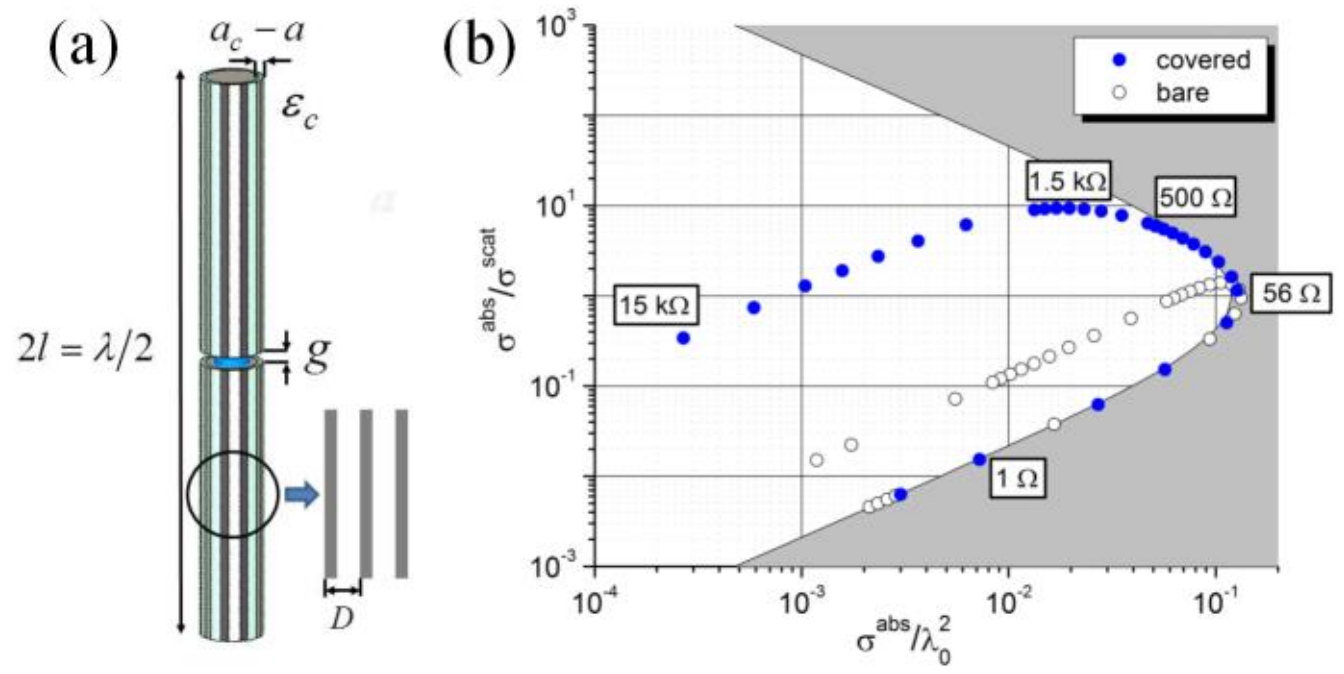

Fig. 6. (a) An half-wavelength dipole antenna covered by a cloaking metasurface designed to reduce its scattering at the resonance frequency; (b) Absorption efficiency of the bare and cloaked dipole for different values of the power absorbed by the antenna.

It is worth noticing that by using more advanced cloaking devices – based on non-linear and parity-time metasurfaces – it is even possible overcoming the fundamental bound and achieving a truly invisible sensor exhibiting zero scattering but still being able to receive the impinging signal [64]-[65]. Such zero-scattering antennas may be used to overcome the self-interference issues in multiple-antenna configurations.

## III. Smart EM Environment with Reduced Scattering from Large Objects

In a typical communication scenario, most if not all the main objects, such as buildings, vehicles, and infrastructures, are by far electrically large compared to the operating wavelength of the EM field supporting the communication. In the past generations (up to 4G), thanks to the longer operative wavelength of the EM carriers, the communication performances were not affected significantly in such complex and crowded environments. On the contrary, with the use of ever higher microwave frequencies, if not millimeter waves, the future generations, i.e., 5G and beyond, will be based on much shorter wavelengths that may rapidly attenuate in a complex environment where long propagation distances must be covered, and multiple scattering takes place. The wireless performances are thus affected by the Line-Of-Sight (LOS) conditions between the transmitter and the receiver, that, when preserved, guarantees the highest communication rate and reliability. In this framework, the smart EM environment can be designed to integrate advanced strategies to avoid the undesired presence of electrically large objects between the receiver and the transmitter.

We recently proposed an electrically large carpet cloak for restoring the transmission beyond an electrically large obstacle as if it were not present [65]-[69]. Its operative principle is based on the EM signal tunneling between two, or more, sets of antennas properly interconnected among them to realize a seamless continuity of the wireless EM environment. To describe its application in a realistic scenario, in Fig. 7(a), we consider the case where the LOS wireless connectivity between two base stations is interrupted by the presence of a railroad tunnel. The proposed antenna-based carpet cloak can be used for by-passing the region of space where the large object is present, though a “capture-and-forward” strategy is realized by its specific design. Indeed, the proposed carpet cloak consists of at least two arrays of antennas that cover the entire device. The elements of the arrays are properly interconnected for realizing a complete passive structure, but with embedded smartness in restoring the wireless connectivity. The smartness is underneath the device surface: in Fig. 7(b) we show how the antennas are interconnected among them in case of an electrically large triangular bump on a reflective ground plane. It is covered by two antenna arrays, *i.e.*, Array A and Array B. The goal of the antenna-based carpet cloak is to restore the reflection from the ground plane as if the bump were not present, obtaining a transmission angle φ always equal to the incident angle θ, realizing a self-adaptive behaviour to the environmental conditions. This is possible thanks to the properly designed passive network. The network has been inspired by both the transmission-line cloaking concept [70], where the incident energy is guided through lines from one side to the other of the cloaked object, and Van-Atta retroreflectors [71], which implement a self-adaptive response exploiting the delay of the phase fronts of the wave illuminating the antenna array. Moreover, the efficient energy transfer from one side to the other side of the electrically large cloak has been made possible by exploiting the enhanced transmission concept enabled by resonant elements connected through a matched transmission line [72]-[76]. In this case, the antennas act as resonant elements and realize a bridge from one side to the other of the triangular bump.

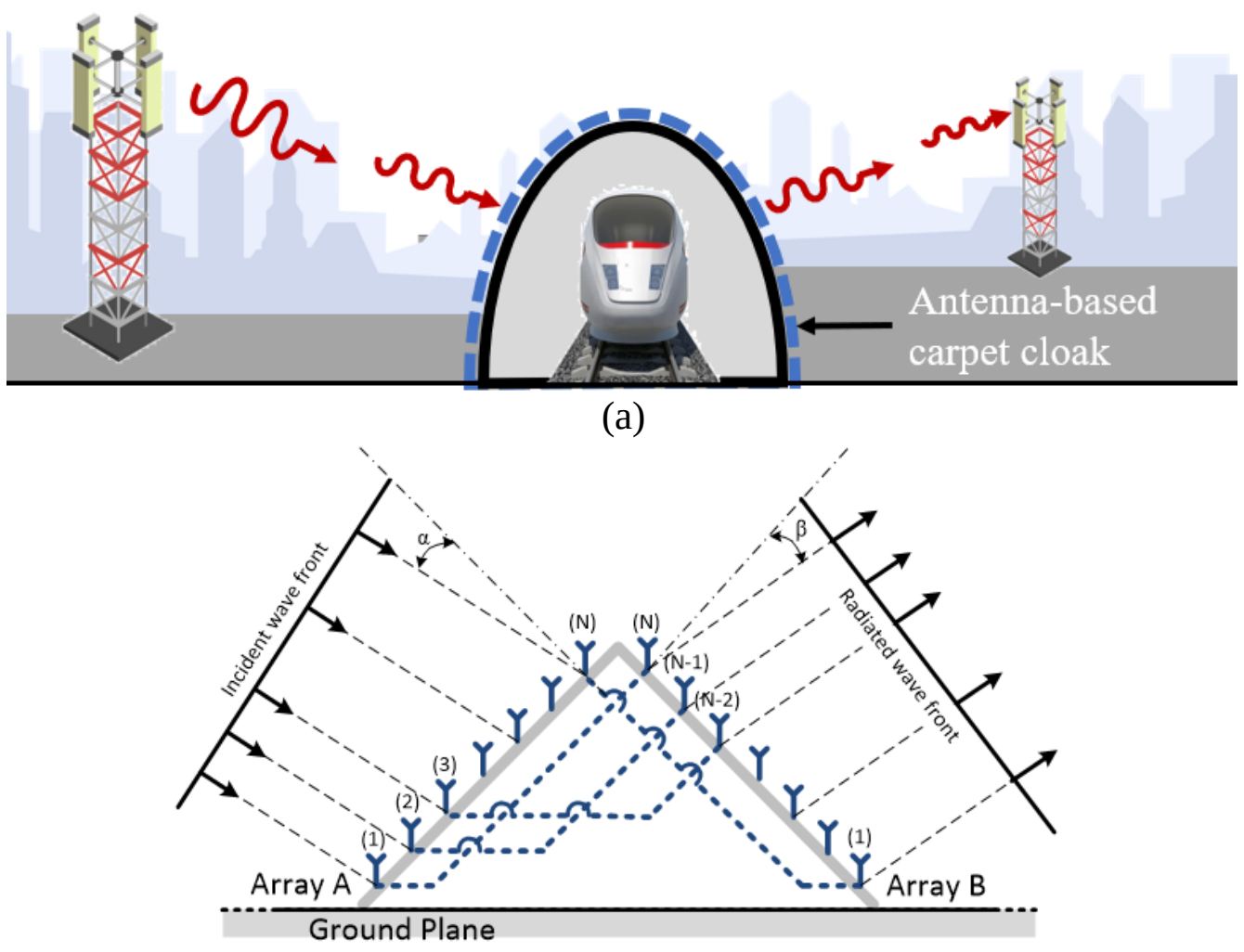

Fig. 7. (a) Pictorial representation of an operative scenario of the antenna-based carpet cloak for electrically large objects: the railroad tunnel forbids the LOS wireless connection between two base stations located at the opposite sites, but the cloak can restore the connectivity by-passing the region of space where the obstacle is present. (b) Operation of the antenna-based carpet device applied to a triangular bump, which restores the reflection from a ground plane for any incidence angle.

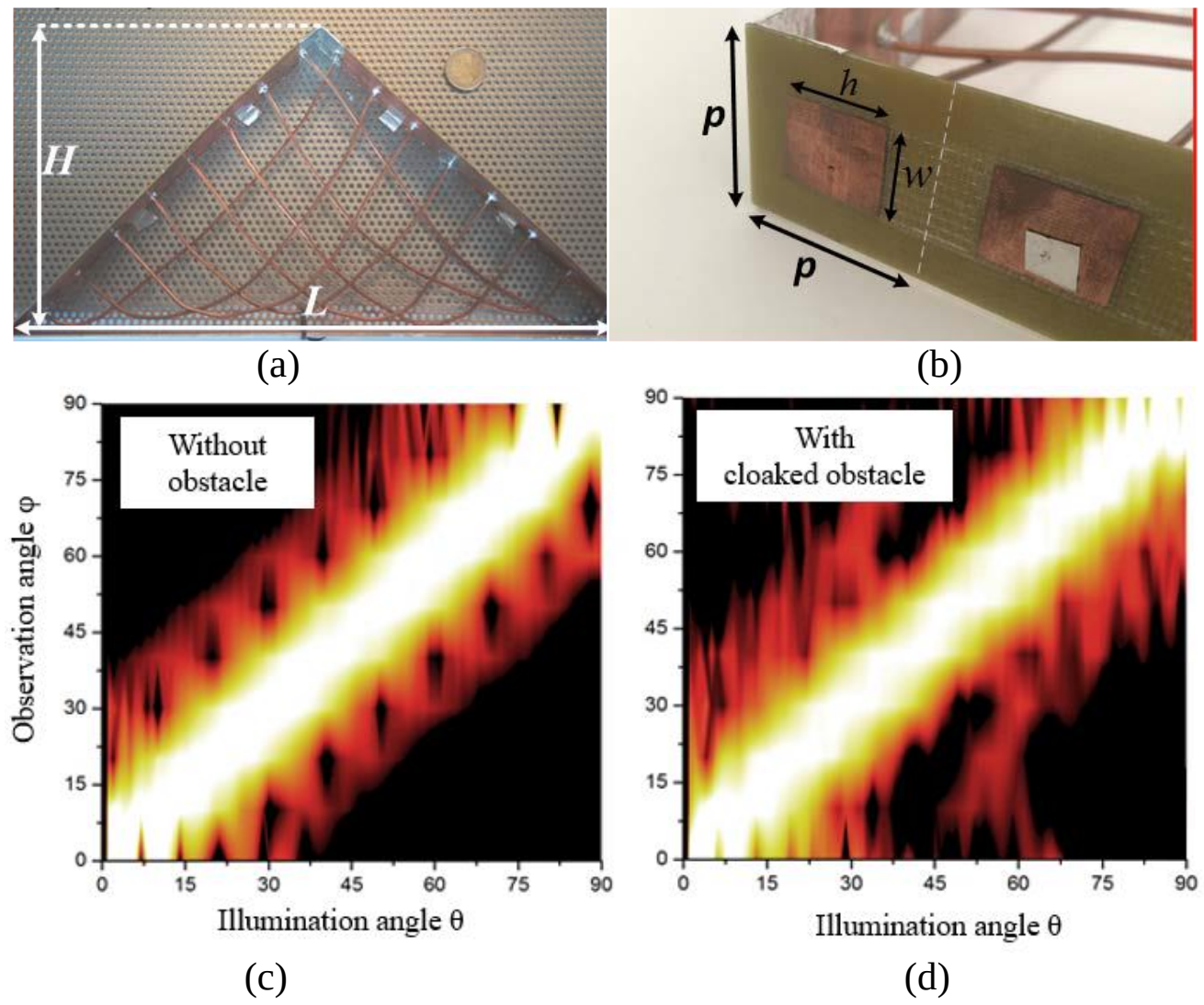

Fig. 8. Realized antenna-based carpet and corresponding results. (a) Side view of the triangular antenna-based carpet device. The antenna elements are connected to each other with a set of 50-ohm coaxial cables of identical length. (b) Zoom on the patch antenna element operating in the sub-6 GHz frequency band. (c)-(d) Normalized electric field amplitude as a function of the incidence and observation angles for the cases: (c) ground plane without any obstacle; (d) in presence of an obstacle covered by the antenna-based carpet cloak.

In [68], we demonstrated experimentally the operativity of the antenna-based carpet cloak covering a metallic electrically large triangular bump of dimensions 2.83 λ x5.6 λ with 14 patch antennas operating at 5 GHz interconnected with a VanAtta-like configuration, as shown in Fig. 8(a)-(b). The numerical and experimental verifications showed that the proposed device was able not only to restore the phase fronts on the other side of the obstacle, but also to operate as an “illumination follower”. Indeed, as shown in Fig. 8(c)-(d), the device has been illuminated at the operative frequency with different illumination angles, ranging from the side illumination, i.e., $\theta = 0°$, to the normal illumination, $\theta = 90°$. For each angular direction, the amplitudes of the normalized electric field for the cases of ground plane “without obstacle” and “with cloaked obstacle,” have been recorded over a similar angular range of the observation angle φ. As expected, in the absence of the obstacle, the maximum of the reflected field is centered around the specular reflection angle and, interestingly, a behavior very similar is achieved when the bump is covered by the antenna-based carpet device.

Complex wireless communication environments can surely benefit from the passive smartness of the proposed cloak for an electrically large object, mainly for avoiding multiple scattering effects and effective longer propagation paths for microwave and millimeter wavelengths and, when possible, restoring the LOS connectivity. In general, the forwarding performances of the proposed cloak in terms of frequency bandwidth and polarization are given by the antennas on its surface, which can be designed to operate in a narrow, multi or wide frequency band and under the desired polarization state. On the contrary, regardless of the antenna element, the wide operative angular behavior is ensured by the underneath Van Atta-inspired interconnection between elements that make this configuration of particular interest for realizing structures that reflect in a direction directly related to the illumination one. Finally, it is worth mentioning that the antenna-based carpet cloak can be theoretically scaled to any electrical size for adapting its surface to cover any arbitrarily large object in the EM environment.

## IV. Smart EM Environment with Zero Fading from Moving Objects

In telecommunications, one of the most destroying effect limiting the performances of a wireless environment is represented by the *fading* [77], which occurs when attenuation appears unpredictably at undesired space and time coordinates. In the case of the EM environment with fast-moving objects, the fading observed at receiver terminals can be caused by the Doppler shift effect, which emulates an effective multipath propagation [78]: the relative speed of a moving source is different for different objects located at different places, and the scattering contributions by these objects may reach the receiver, where they are combined together in a received signal strongly affected by a time-dependent fading due to the relative motion of the objects in the wireless environment. As an example, let us consider a transmitter moving in a certain direction; the several contributions reaching the receiver would be at slightly different frequencies due to the relative Doppler shift experienced by the EM waves in their own path. In a way like the phase-shifted signals that cancel each other by destructive interference, the frequency-shifted signals interfere and create fading. As a side effect, the Doppler multipath causes jitter and an increased inter-symbolic interference (ISI).

In this context, the Doppler cloak technology has been recently introduced for achieving an effective stationarity of moving objects through a time-modulated scattering response [79]-[82]. The original concept of the Doppler cloak has been proposed exploiting the intermodulation properties of the spatio-temporal modulated metamaterials, which are able to realize a temporal convolution between the illuminating signal and the dynamic permittivity function of the modulated artificial bulk metamaterials [79]. During the propagation within the modulated media, the illuminating wave is coupled to two, or more, modes supported by the system and perform a perfect energy transfer from the original frequency $f$ of the incident signal to the shifted frequency $f \pm \delta f$ of the scattered field. This property has been initially exploited for breaking the reciprocity in antenna systems, *i.e.* having a different realized gain in transmission and reflection, without the use of magnets [83]-[86], but the controllable frequency shift induced by the modulation represents also the key for compensating the Doppler effect and, therefore, vanishing the apparent motion of the objects [80].

In the framework of smart EM environments, let us consider the realistic scenario depicted in Fig. 9 where Doppler cloak technology is applied to a car. Despite it is in motion, the reflected field by the red car is at the same frequency of the source and, obviously, of the city background; on the contrary, the blue car reflects a Doppler red-shifted frequency due to its motion away from the base station introducing the undesired time-dependent variation of the environment.

The metamaterial-based Doppler cloak has been demonstrated to be effective in these cases, but in general, its application is limited by the electrical dimensions and the complexity in the modulation scheme of the bulk material properties. Therefore, in [87] an electrically thin reflective metasurface has been proposed for inducing an artificial

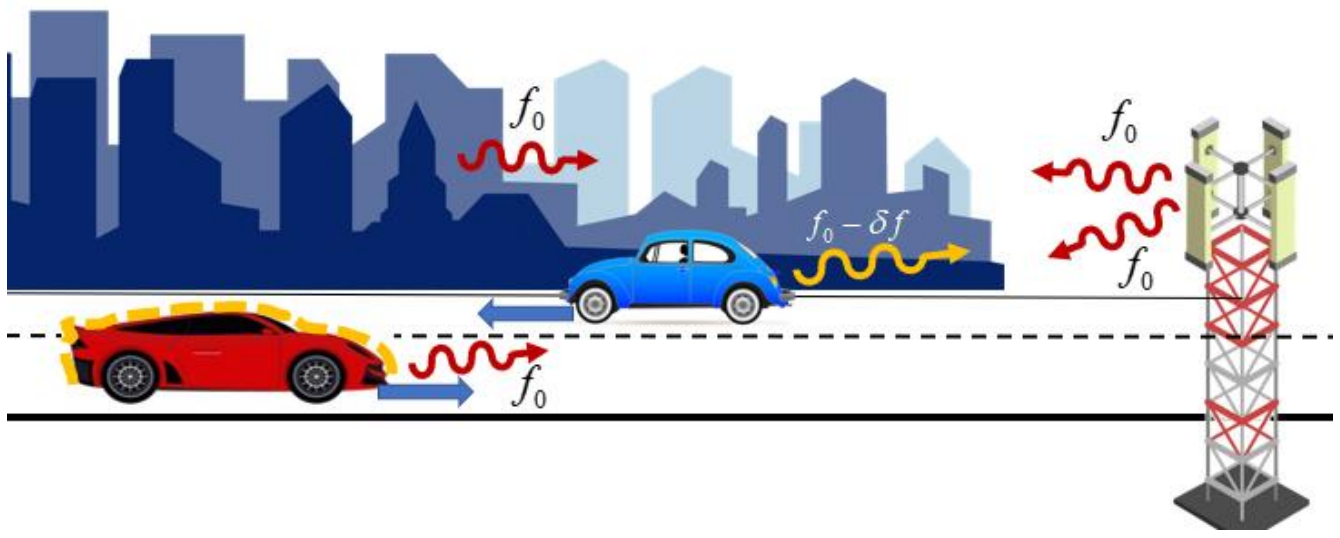

(a)

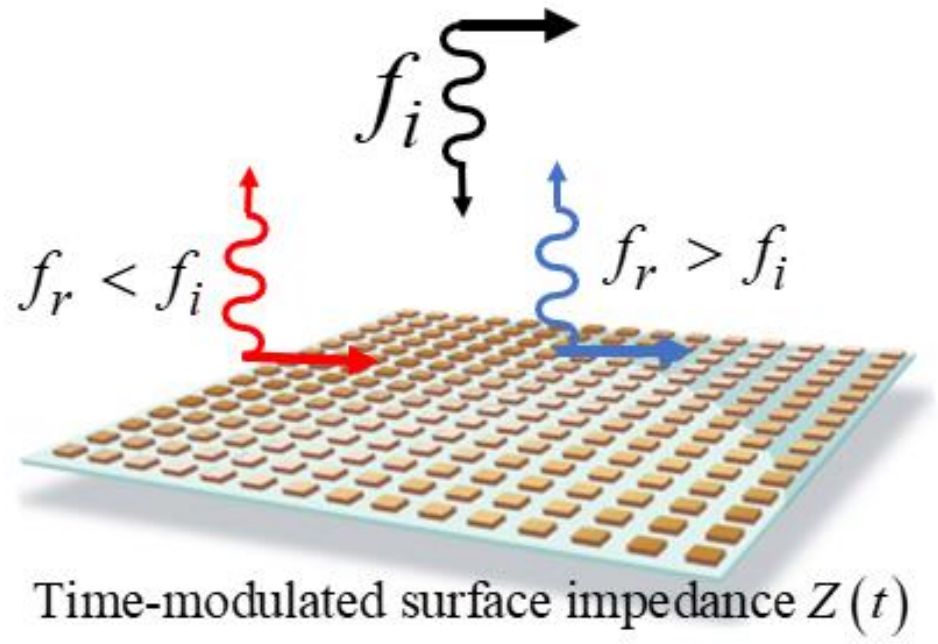

(b)

Fig. 9. (a) Example of the Doppler cloak technology applied to an urban environment: the red car is covered by the Doppler cloak and, despite it is in motion, the reflected field is at the same frequency of the static background; on the contrary, the blue car reflects a Doppler red-shifted frequency due to its motion away from the base station. (b) Graphical representation of a metasurface Doppler cloak, composed by a time-modulated metasurface whose surface impedance varies periodically over time. The illuminating field at frequency $f_0$ is reflected with a higher or lower frequency according to the modulation signal applied to the metasurface, realizing a controllable Doppler effect.

Doppler shift acting on the time-depended surface properties (Fig. 9(b)). The operative principle is based on the single-side frequency translation achieved by a pure phase contribution in reflection imposed by the metasurface. Indeed, according to the signal theory, to shift a spectrum in frequency, the corresponding signal in the time-domain must be multiplied by a time-varying complex exponential $\exp[i\phi(t)]$ [87]-[89]. To emulate it at the EM level, the incident and the reflected field should be related to each other by the following reflection coefficient as a function of time: $\Gamma(t) = e^{j(\delta f \cdot t)}$, where $\delta f$ is the frequency shift imparted by the modulation. Such a reflection coefficient can be easily achieved by a time-varying metasurface exhibiting unitary reflection and time-varying phase. This can be implemented by exploiting the reflection response of a high-impedance surface (HIS) [90], that can exhibit any phase of the reflection coefficient between the two extreme conditions of a perfect electric conductor, whose reflection phase is $\pm 180°$, to perfect magnetic conductor, whose reflection phase is zero. The dynamic tunability of the reflection phase at the operating frequency can be achieved by loading the surface with a set of reverse-biased varactors, as shown in Fig. 10(a). When the modulation signal is applied to the varactors, the response of the HIS is dynamically varied to make the surface able to exhibit the full phase coverage from -180° to +180° over a time modulation period. This emulates the motion of the surface in the normal direction to the surface itself and induces an artificial Doppler shift to the reflected field. In Fig. 10(b)-(c), we report the normalized spectrum of the reflected field in the case of up- and down-converting modulation, respectively.

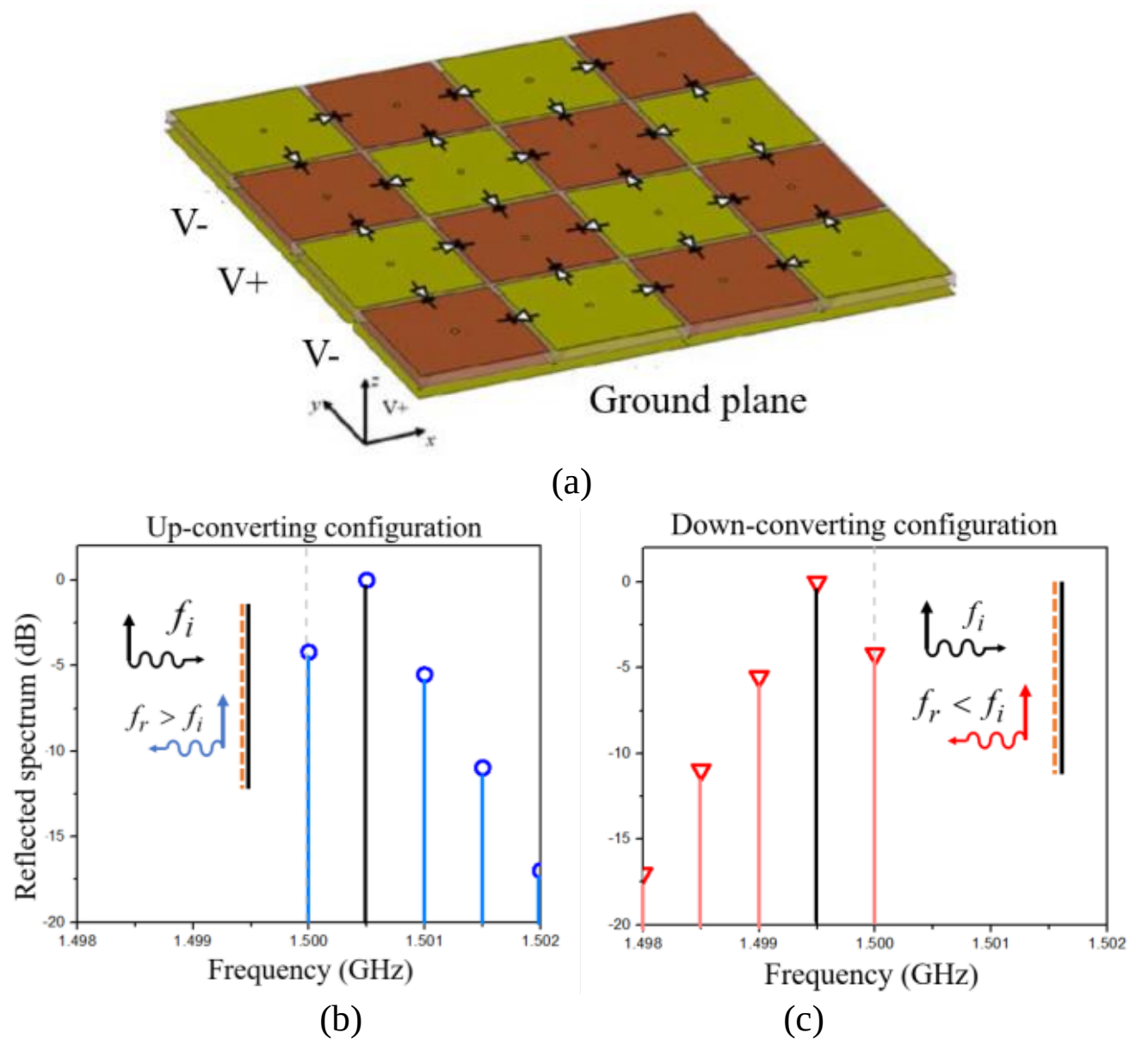

Fig. 10. (a) Metasurface Doppler cloak technology implemented through a high-impedance surface whose response is temporally modulated with a set of varactors. (b) – (c) Normalized spectrums of the reflected signals from the metasurface Doppler cloak in case of (b) up-converting configuration and (c) down-converting configuration.

## V. Smart EM Environment with Dynamic Scattering Control

To satisfy the severe KPIs requirements of future wireless communications, dynamic transformation of the radiator characteristics depending on the environment properties will be an enabling key feature. A new generation of antenna systems equipped with reconfigurable, tunable, time-modulated, or even programmable properties, in fact, is expected to lead the future of the next generation of wireless communications.

In this regard, advanced metasurfaces can be used to surrround wire antennas, enabling an on demand engineering of the electrical, scattering, and radiating properties, adapting the antenna behavior to the environment conditions and, thus, dramatically increasing the degrees of freedom available to the antenna designer (Fig. 11). In particular, in the following, we focus on advanced metasurfaces whose unit-cells are loaded with electronic elements that can be controlled to achieve the desired EM functionality.

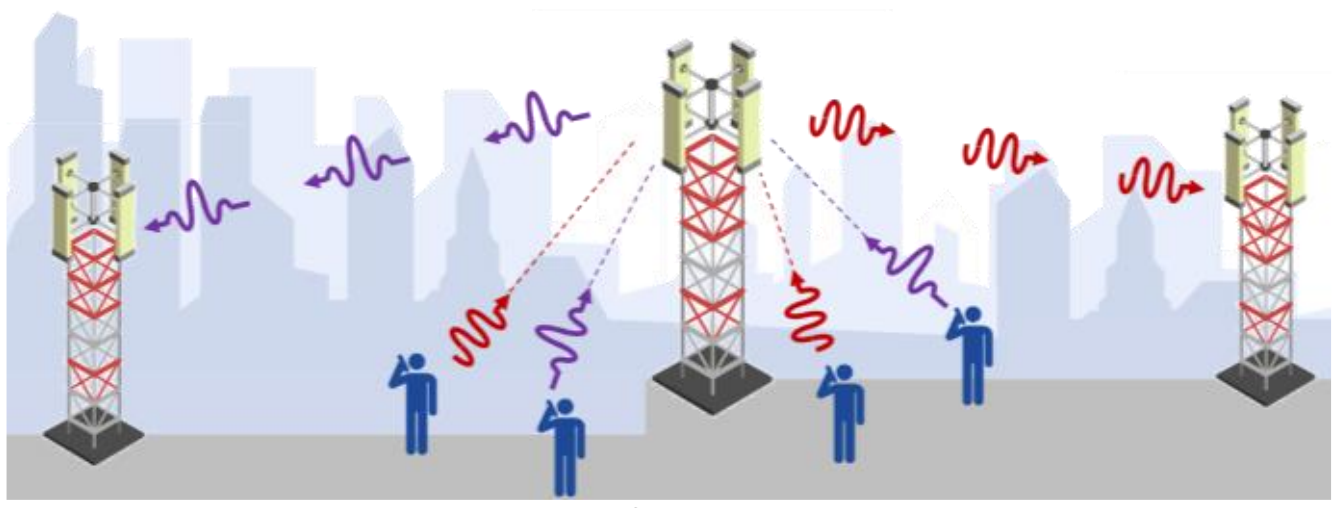

Fig. 11. Pictorial representation of a cognitive antenna system where the antenna radiation properties are dynamically reconfigured depending on the environment characteristics. Depending on the properties of the received signals (e.g., frequency, polarization, power level, temporal waveform, etc.) the base station reconfigures itself to adapt to the environment.

### A. *Power-selective cloaking*

As a first example, we report the design of power-dependent cloaking metasurfaces for wire antenna arrays [65],[91][95]. In the previous sections, we have shown how properly engineered homogeneous metasurface can be used to suppress the scattering signature of wired antennas at a specific frequency of operation [22]. This functionality can be further enriched by loading the cloaking metasurface with PIN diodes for designing non-reciprocal antenna systems whose scattering and radiating properties depend on the power level of the received/transmitted signal. By increasing the cloaking metasurface design complexity, it is possible to use diode pairs to make the patterned geometry of the metasurface varying accordingly to the power level of the incoming EM wave, thus, introducing a dynamic EM response of the structure as a function of the incoming power-level [92].

In Fig. 12(a), the case of a dipole antenna coated by an inductive cloaking metasurface made of conventional vertical metallic strips is reported [93]. These strips are interleaved by a sub-array of three truncated strips. The gaps between them are loaded with a PIN diode. Thanks to the non-linearity of the diode, the truncated strips are almost short-circuited in presence of a high-power level signal (HP), whilst the strips are open-circuited for a low-power-level signal (LP) [94]. The inductive metasurface periodicity is thus modified accordingly to the signal power level, as well as its EM characteristics [95]. As shown in Fig. 12(b), the wire antenna coated by such a metasurface exhibits extremely low scattering at the resonant frequency $f_0$ = 3 GHz in presence of a LP signal (i.e., its visibility level is dramatically reduced), whilst its scattering properties and visibility level are restored for a HP signal. Here, the surface impedance value of the metasurface is modified thanks to the dynamic periodicity variation of the strips, and the cloaking resonance is shifted towards lower frequencies (2.5 GHz).

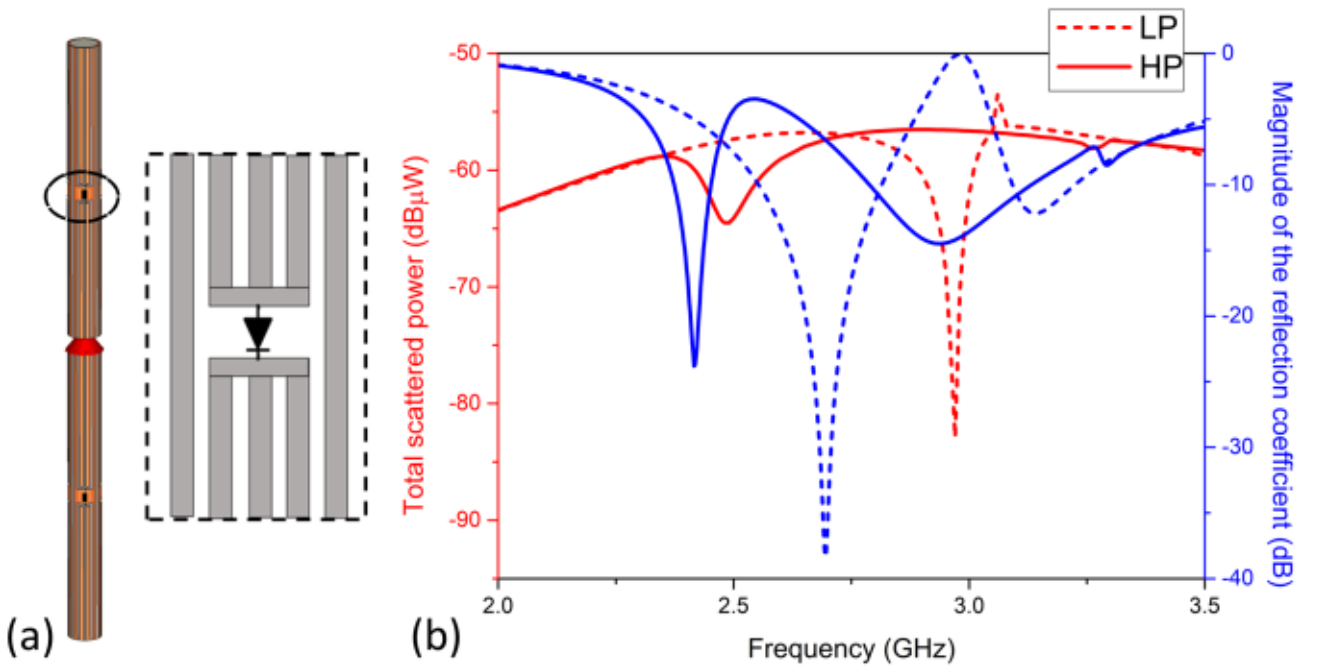

Fig. 12. Dipole antenna covered by the non-linear cloaking metasurface. In the inset, detail of the metasurface made of vertical continuous strips interleaved by truncated ones and loaded with a diode pair. (b) Total scattering signature (red curves) and magnitude of the reflection coefficient (blue curves) of a half-wavelength dipole antenna coated by an inductive power-dependent mantle cloak, for HP and LP signals. In the inset, detail of the metasurface made of vertical continuous strips interleaved by truncated ones and loaded with a diode pair.

It is worth noticing that also the antenna reflection coefficient is modified accordingly. For a LP signal, while the antenna visibility is highly reduced at $f_0$, the antenna is mismatched at its working frequency. This effect is consistent with the absorption and efficiency limits of antennas and the optical theorem [43], and it is an indication of the dramatic scattering minimization. Indeed, for a HP signal, the matching characteristic of the antenna at $f_0$ is restored and the antenna can radiate efficiently.

The mentioned non-linearity of the cloaking effect can be used to design an unconventional array system able to transmit and scan selectively the environment, whilst receiving the scattering signal from all directions of space. In Fig. 13, the scenario of a 3x3 dipole antenna array is reported. As well known, by engineering the array factor of the system the radiation pattern of the array can be made very sharp. However, due to the mutual coupling interaction between the radiating elements, the embedded array factor of the single element is not omnidirectional, even though the individual radiating element is omnidirectional when operating in free space, resulting in limited scanning ranges and undesired scan loss effects.

Indeed, by equipping the peripheral radiating elements of the array with the non-linear cloaking device, this limitation can be circumvented. In fact, in the transmitting mode, a high-power level signal impinges onto the cloaking metasurfaces coating the external elements, short-circuiting the truncated metasurface strips and, thus, switching OFF the cloaking functionality at $f_0$. The array exhibits, thus, a sharp directional pattern and can scan selectively the environment. However, in the receiving mode, the LP signal scattered by the environment is received and the cloaking effect is switched ON, reducing the visibility of the peripheral elements of the array. Therefore, the central radiating element of the array is characterized by an omnidirectional pattern and can efficiently receive from any direction.

The discussed radiating system is only an example of the possibilities offered by this design approach for conceiving antenna systems able to adjust their operation to the environment power-level. Further radiating devices with enriched functionalities can be envisioned exploiting non-linear mantle cloaks [92].

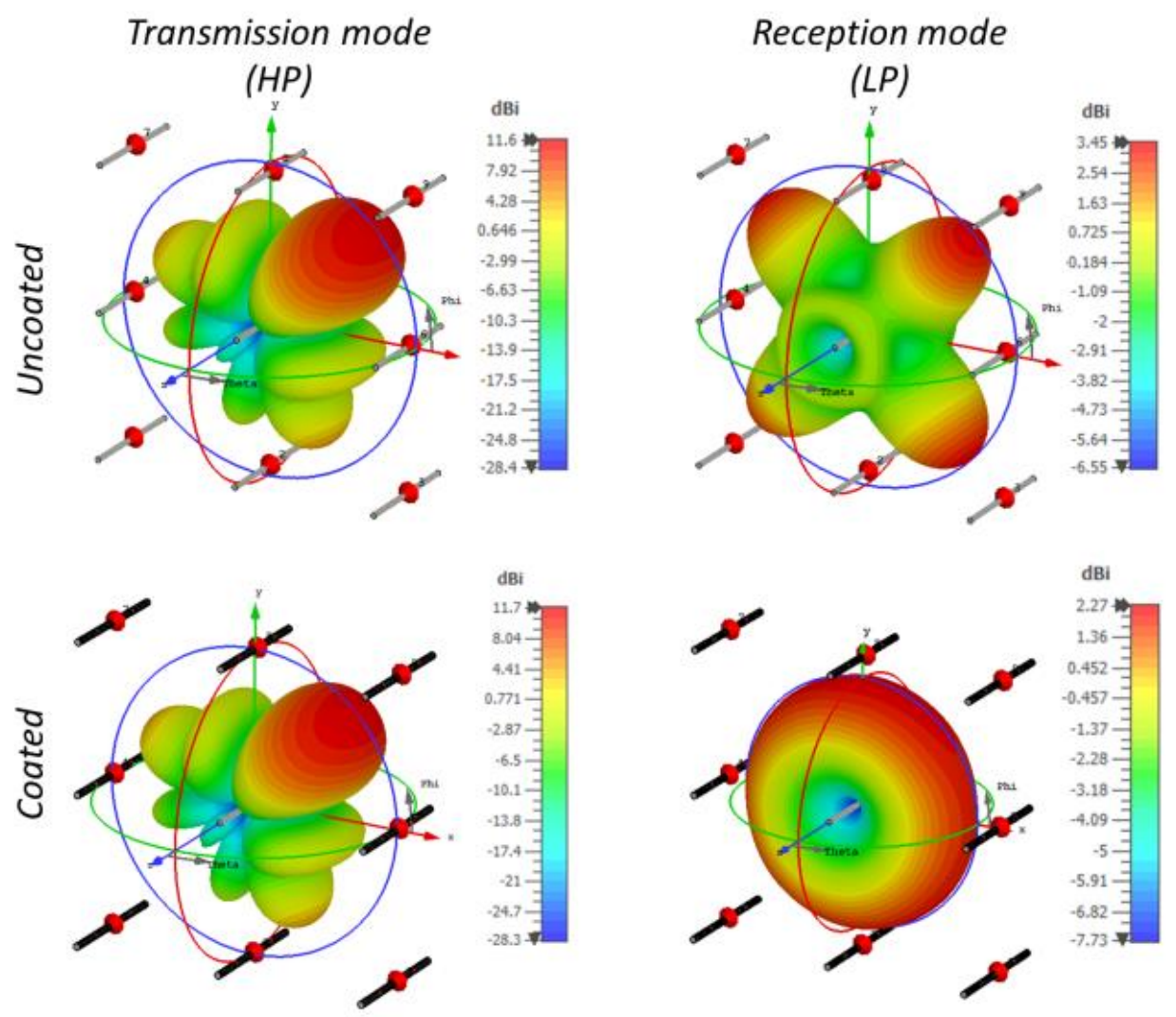

Fig. 13. Radiation patterns of the phased array of dipole antennas (left) and embedded central element pattern (right) with uncoated/coated peripheral elements for HP/LP signals excitations.

### B. *Waveform-selective cloaking*

The scattering and sensing capabilities of an antenna system can be further manipulated by equipping the coating metasurface with a more advanced electronic circuit, to address the strong demand of developing ever more advanced wireless communication technologies and enhance the quality of telecommunication services. Recently, it has been shown that circuit-based absorbing metasurfaces can be used to reduce interferences coming from different signals at the same frequencies by discriminating the temporal waveform of the incoming signal [97],[98]. This new approach gives us a completely new degree of freedom in the antenna design, allowing to control EM waves depending not only on the frequency, polarization, power level, etc. but also on the temporal waveform characteristics.

As a relevant example, in the following, we report the case of an antenna made invisible to a very short pulsed signal, such as the one coming from a detecting radar, whilst the visibility level of the antenna is restored in presence of a signal with a larger pulse width, such as the one coming from a radio base station [99],[100]. we show that in the latter case the antenna can efficiently receive/transmit, and the EM functionalities are restored. In Fig. 14(a) the principle of the antenna design is reported. A half-wavelength dipole antenna working at $f_0$ = 3 GHz is coated by a metasurface made of meandered unit cells. The meander of each unit-cell is loaded by the lumped element circuited reported in Fig. 14(b).

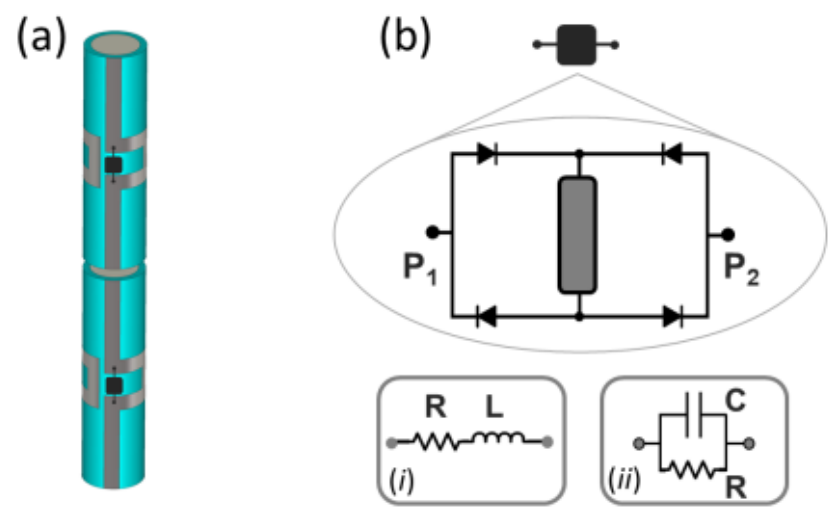

Fig. 14. (a) Sketch of the dipole antenna coated by the waveform-selective metasurface. (b) Schematic of the loading waveform-selective circuit composed by a full-bridge diode rectifier and: (*i*) a RL series, (*ii*) a shunt RC.

In particular, the circuit is composed of a full-bridge rectifier and a series load made by, in this case, a series between an inductor L and a resistor R. The peculiar waveform dependent response of the metasurface is due to the specific transient response of the circuit: in presence of an external source at the input port of the circuit, the incoming signal is rectified and most of the energy is converted to zero frequency. Thus, in the beginning, the current flow along the circuit is inhibited due to the presence of the inductor and its opposing electromotive force. However, after some time, this opposing effect is reduced, and currents are allowed to flow towards the output port [100],[101].

Therefore, for a signal characterized by a very short pulse width Δt compared to the time constant of the circuit $\tau = L/R$ (i.e., a pulsed waveform - PW) current flow is not allowed, whilst for a signal with $\Delta t >> \tau$ (i.e., a continuous waveform - CW) strong currents, almost short-circuiting the input and output port of the circuit, can flow. It is worth noticing that the opposite behaviour can be achieved by loading the diode bridge rectifier with a parallel combination between a resistor and a capacitor [97],[102].

The scattering behaviour of the antenna once coated with the deigned circuit-loaded metasurface is reported in Fig. 15(a) and compared to the uncoated scenario. In presence of a PW, the scattering signature of the antenna is minimized. Here, the waveform-selective circuit behaves almost as an open-circuit and the metasurface behaves almost as if the circuit load were not present. The patterned meander geometry is, thus, not affected by the presence of the circuit and, since the unloaded metasurface is engineered to suppress the antenna scattering signature, the device visibility is strongly reduced. On the contrary, when the antenna receives a CW, the meander of the unit-cell is short-circuited due to the response of the circuit. The metasurface unit-cell is, thus, dynamically transformed from meander-like to a vertical strip and its EM response is modified accordingly. In particular, this geometry variation results in a cloaking resonance shift. Hence, the scattering signature of the antenna at its working frequency is restored and its visibility level recovered, as confirmed also by the radiation pattern at $f_0$ when compared to the uncoated case.

Finally, in Fig. 15(b) the magnitude of the reflection coefficient of the antenna is reported. In the uncoated antenna scenario, the conventional resonance can be recognized. Once the antenna is coated by the loaded metasurface, two different behaviours can be appreciated depending on the temporal waveform of the signal. For a PW a strong mismatching is present at the antenna input port when loaded to a conventional 50 Ohm load. As mentioned before, this is an effect due to the cloaking behaviour that is turned on. In this case, the antenna functionalities are inhibited. For a CW, instead, the matching characteristics are restored, and the antenna can receive/transmit efficiently since the cloaking effect is turned off.

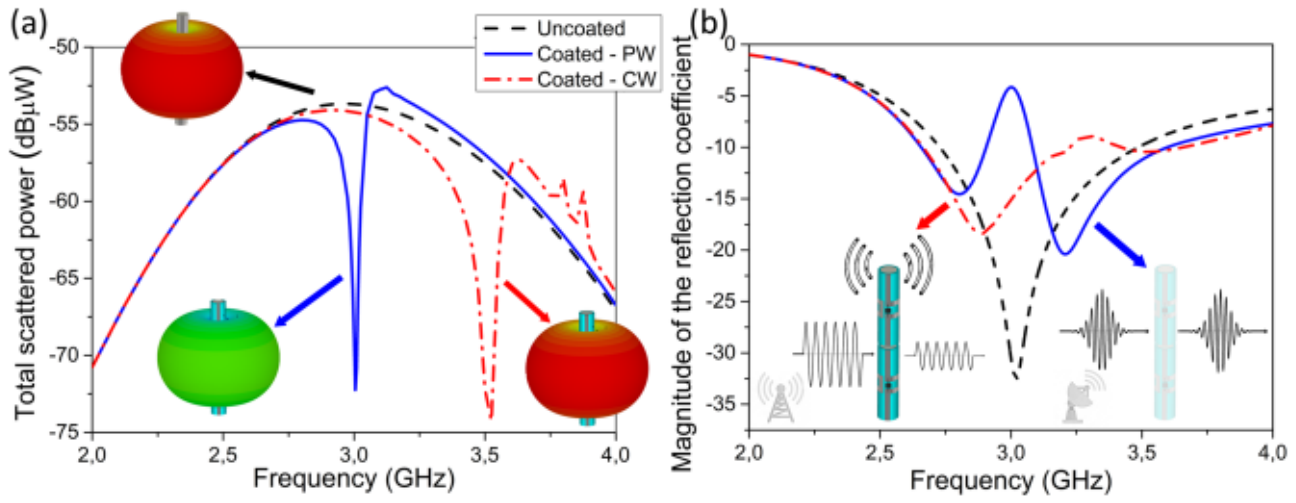

Fig. 15. (a) Total scattering signature of the antenna in the uncoated and coated scenarios under PW or CW excitations. In the insets, realized gain radiation diagrams of the antenna in the different cases evaluated at $f_0 = 3$ GHz. (b) Magnitude of the reflection coefficient of the antenna in the uncoated and coated scenarios under PW or CW excitations. In the insets, sketch of the corresponding antenna's visibility level.

## VI. Surfaces with Controllable Reflection and Refraction for Smart EM Environment

As mentioned before, when increasing the frequency, the LoS link is ever more affected by the fast free-space attenuation and by the presence of objects in the environment. In beyond 5G communications, thus, the environment itself must play an active role to relax the computation efforts: walls, doors, windows, external surfaces of objects and devices must evolve from mere obstacles to smart EM relay nodes. The use of such smart surfaces would also allow ultra-fast forwarding and routing with a dramatic improvement of the overall system performances. Ideally, as shown in Fig. 16, a smart surface equipped with a local control and a low computation module enables sensing of the the environment (e.g. estimation of the direction of arrival), positioning awareness (e.g. absolute and relative position of a node) and adaptive connectivity (e.g. permanent restoration of LoS, overcoming localized coverage holes).

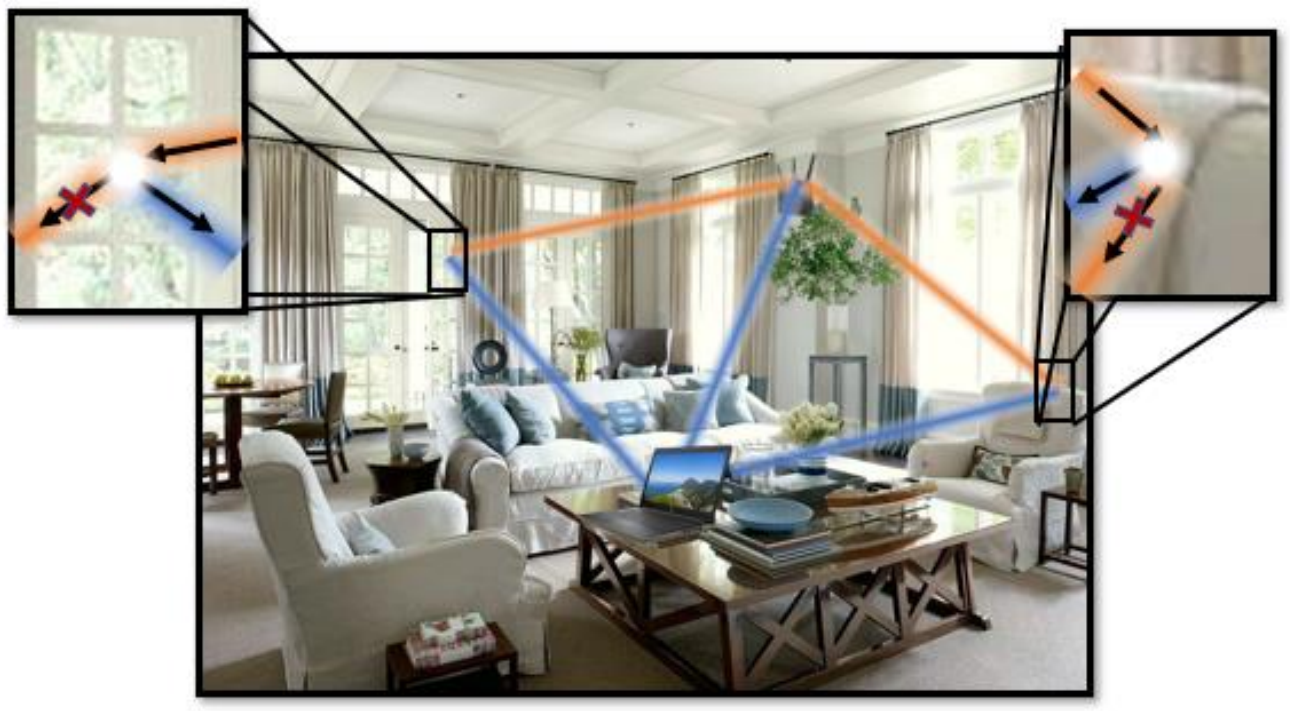

Fig. 16. Illustrative sketch of a smart environment in which the surfaces of the objects become ultra-fast relay nodes, making the environment playing an active role in the communication link.

A possible approach for achieving controllable reflection and refraction by a surface consists in the use of reconfigurable meta-gratings [103]-[104]. These structures are composed by a periodic lattice of electrically or magnetically polarizable elements enabling advanced diffraction engineering. As an example, we have recently proposed a reconfigurable beam steering meta-grating loaded by varactor diodes, where dynamic diffraction patterns can be obtained by applying different bias voltages to the diodes, resulting in different values of the capacitances connecting the physical parts of the unit-cells of the meta-grating.

The structure shown in Fig. 17 (a) has been proposed in [105]. It consists of a dielectric layer sandwiched by two arrays of conducting loaded strips which is illuminated by a TE-polarized plane wave with the incident angle $\theta_{in}$. The substrate and the arrays are infinite in the *x*- and *y*- directions. In this scenario, the total electric field on the upper and lower medium of the structure can be evaluated as a superposition of infinite Floquet modes by generalizing the analytical model introduced in [106], and, it can be demonstrated [105], that the amplitude of the diffracted waves becomes a function of the load impedances. Therefore, any desired refracted or reflected waves can be realized by adjusting the power coupled to each Floquet mode.

To verify this concept, a metagrating that can excite three transmitted and reflected Floquet modes (m=0,±1) has been designed. The diffracted angle $\theta_{out}$ for m= ±1 is equal to arcsin($\lambda_0/\Lambda$), being $\Lambda$ the periodicity of the structure in the *y*-direction. Therefore, once the wavelength is fixed, the desired $\theta_{out}$ can be obtained by adjusting the periodicity. Here, $\theta_{out}$ was fixed to $60^0$, corresponding to a periodicity of 1.55 $\lambda_0$. Moreover, four strips in each unit cell (N=4) were considered. Therefore, the evaluation of eight load capacitance for the upper and lower arrays was required. Genetic algorithm has been used to optimize the required load impedances.

In particular, the electric field distribution for the case of anomalous reflection with $\theta_{out}=60^0$ at the operation frequency of 3.5GHz is shown in Fig. 17(b). It can be seen that the EM wave reflects with the desired angle. From Fig. 17(c), it can be concluded that the efficiency of the reflection with the desired Floquet mode is 97% and just a small portion of the incident wave propagates on other modes.

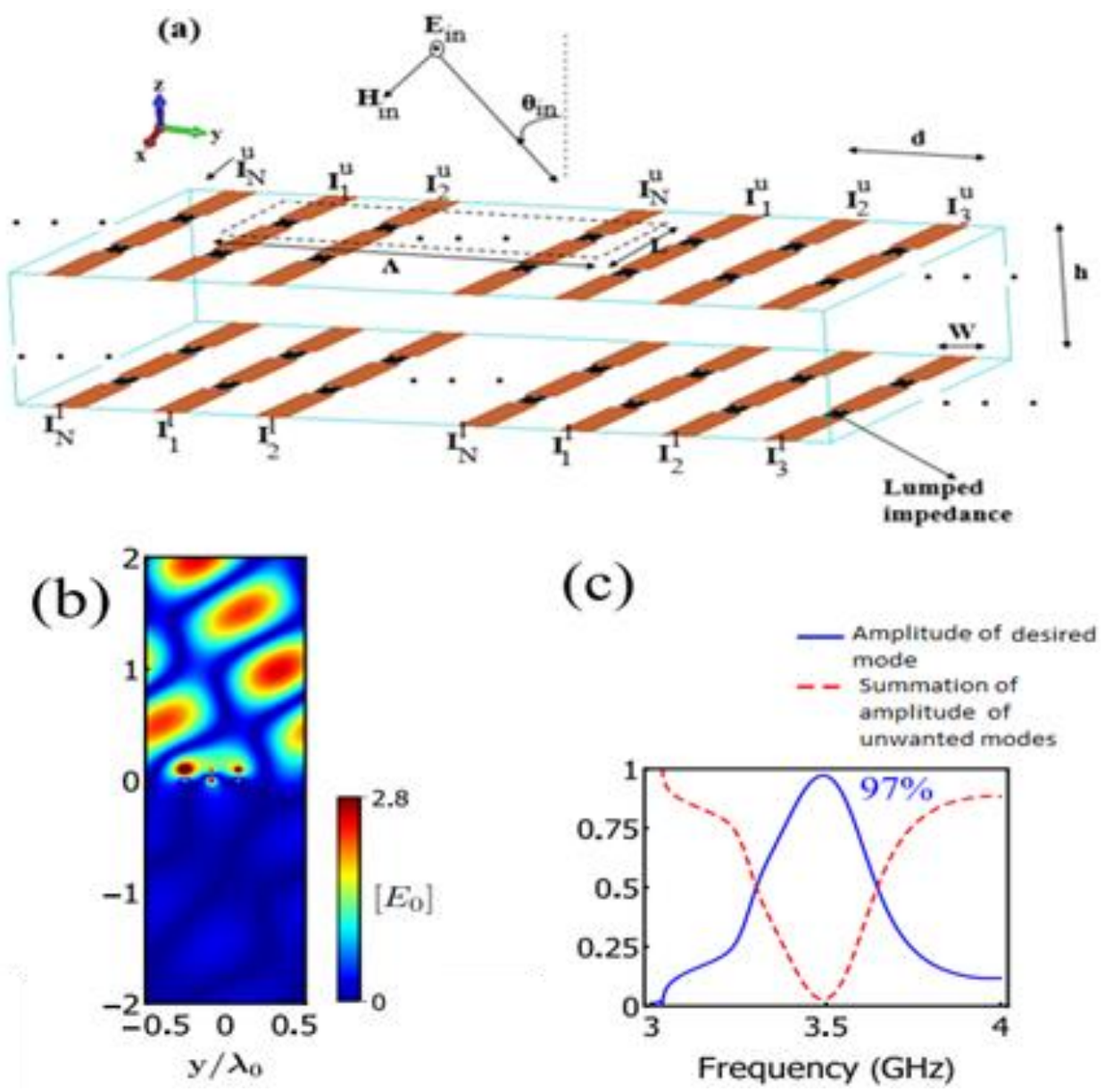

Fig. 17. (a) Structure of proposed metagrating; (b) Electric field distribution for anomalous reflection. (c) Amplitude of the desired mode and summation of unwanted modes.

To implement reconfigurability with the proposed static structure, the mategrating strips were loaded with varactor diodes with tunable capacitance with respect to the applied bias voltages. Therefore, by adjusting the load capacitances, desired scattering manipulation of the EM waves can be obtained with a fixed structure. In particular, each strip was connected to a bias voltage through a resistance and a conducting via, and a varactor diode with capacitance values varying from 0.6 to 0.055 pF was used [105].

Fig. 18 (a)-(b) depicts the performances in terms of anomalous reflection and refraction of the reconfigurable meta-grating through the plot of the electric field distribution for the same case of anomalous reflection and refraction of the static meta-grating, respectively. It is worth noticing that, the price to pay to achieve reconfigurability is a reduction of the system efficiency that drops almost 20% in comparison to the static one. This drawback, however, is mainly due to the power dissipated in the varactor and bias resistances. It is possible to overcome this problem by using MEMS varactors that provide more linearity and higher quality factor [107]. The additional important feature of the proposed dynamic meta-grating is its inherent reconfigurability, enabled by the electronic control of the loading capacitances.

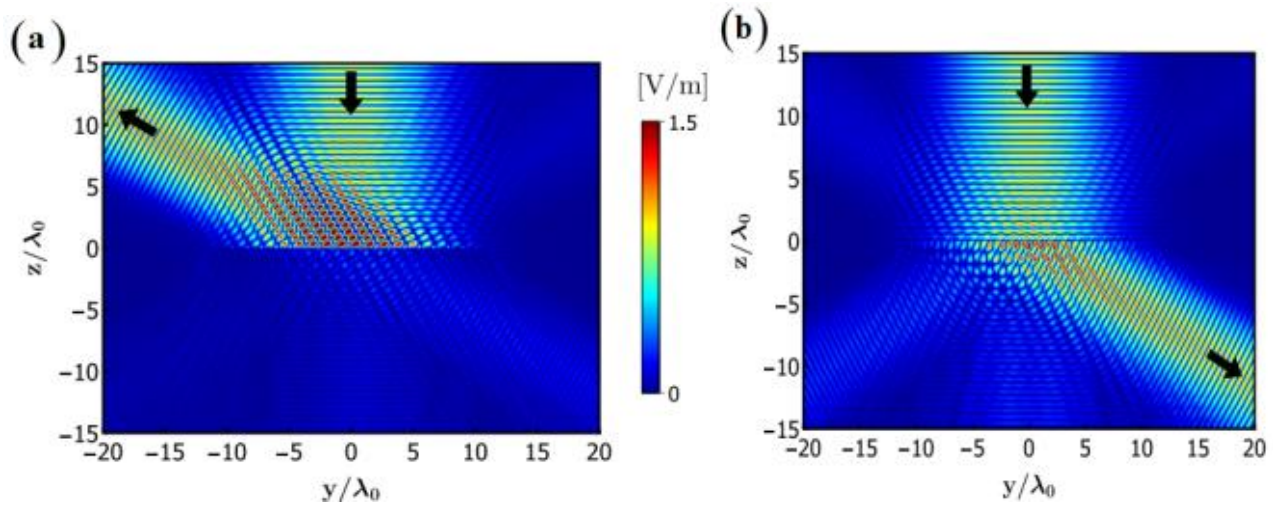

Fig. 18. Electric field distribution for the reconfigurable metagrating for anomalous (a) reflection and (b) refraction. Comparison of the radiation pattern for static and reconfigurable structure for anomalous (c) reflection and (d) refraction.

## VII. Conclusions

In this paper, we have presented and discussed the potentialities of electromagnetic metasurfaces for overcoming the main detrimental effects introduced by the environment in the operation of current and future communication systems. With several relevant examples – such as the blockage effects occurring in overcrowded communication platforms, the no-LOS communication due to obstacles, and the time-varying fading due to fast-moving scatterers – we have shown how metasurfaces may make the external environment an active element of the next-generation communication systems, *i.e.*, may allow conceiving a smart electromagnetic environment. Besides, we have also discussed how the degrees of freedom offered by metasurfaces allow relocating to the physical layer the challenges of beyond-5G communication systems, minimizing hardware virtualization, computational resources, and elaboration time. We expect these findings may confirm the suitability of metasurfaces as a promising solution for implementing intelligent surfaces and further boost the research efforts on these topics.

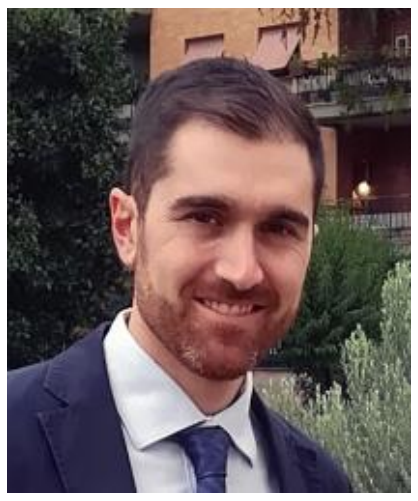

**MIRKO BARBUTO** (S'12–M'15-SM'19) was born in Rome, Italy, on April 26, 1986. He received the B.S., M.S. and Ph.D. degrees from "Roma Tre" University, Rome, Italy, in 2008, 2010 and 2015, respectively.

Since September 2013, he is with the "Niccolò Cusano" University, Rome, Italy, where he currently serves as an Associate Professor of electromagnetic field theory, as the Director of the Applied Electromagnetic Laboratory, and as a member of the Doctoral Board in Industrial and Civil Engineering.

His main research interests are in the framework of applied electromagnetics, with an emphasis on antennas and components at RF and microwaves, cloaking devices for radiating systems, metamaterials and metasurfaces, electromagnetic structures loaded with non-linear or non-foster circuits, topological properties of vortex fields, and smart antennas for GNSS and communication technology.

He serves as Associate Editor for IEEE AWPL (since 2019) and, for this role, he has been awarded for the exceptional performance from 1 January to 31 December 2020. He is a member of the Editorial Board of the Radioengineering Journal (since 2019), of the Technical Program Committee of the International Congress on Artificial Materials for Novel Wave Phenomena (since 2017), and of secretarial office of the International Association METAMORPHOSE VI (the Virtual Institute for Artificial Electromagnetic Materials and Metamaterials). He serves as a Technical Reviewer of the major international conferences and journals related to electromagnetic field theory and metamaterials and he has been the recipient of the Outstanding Reviewers Awards assigned by the Editorial Board of the IEEE Transactions on Antennas & Propagation (for six consecutive years, 2015-2020) and by the Editorial Board of the IEEE Antennas and Wireless Propagation Letters (for three consecutive years, 2017-2019). In 2017 he has been selected as one of the Best Reviewers by the Editorial Board of Radioengineering Journal. He has been Technical Program Coordinator (Track "Electromagnetics and Materials") for the 2016 IEEE Antennas and Propagation Symposium and he served as Guest Editor of three special issues on metamaterials and metasurfaces.

Since 2015, he is the Proceeding Editor for the annual International Congress on Engineered Material Platforms for Novel Wave Phenomena – Metamaterials and, in 2019, he has been appointed as General Chair of the 39th EUPROMETA doctoral school on metamaterials held in Rome, Italy.

Prof. Barbuto is currently a member of the Italian Society on Electromagnetics (SIEM), of the National Inter-University Consortium for Telecommunications (CNIT), of the Virtual Institute for Artificial Electromagnetic Materials and Metamaterials (Metamorphose VI AISBL), and of the Institute of Electrical and Electronics Engineers (IEEE). Currently, he is the author of more than 90 papers in international journals and conference proceedings.

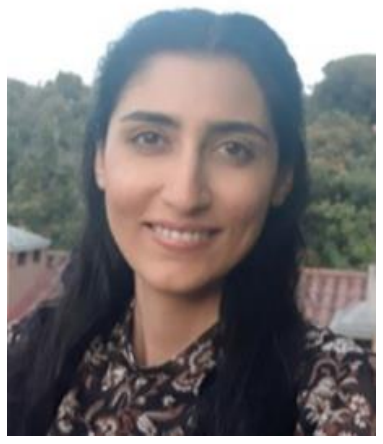

**ZAHRA HAMZAVI-ZARGHANI** was born in Shiraz, Iran. She received the B.Sc. degree in electrical engineering from Shiraz University, Shiraz, Iran in 2010, and the M.Sc. degree in electrical engineering from Tarbiat Modares University, Tehran, Iran, in 2014. She received her joint Ph.D. degree from Shiraz University, Shiraz, Iran and Politecnico di Torino, Turin, Italy in 2020. Currently she is a postdoctoral researcher at Roma Tre University, Rome, Italy, where she is working on electromagnetic and acoustic metasurfaces. Her research interests include metamaterial and metasurfaces, Electromagnetics and Acoustics, imaging, scattering manipulation and cloaking, transmitarray and reflectarray antennas, graphene, and tunable applications.

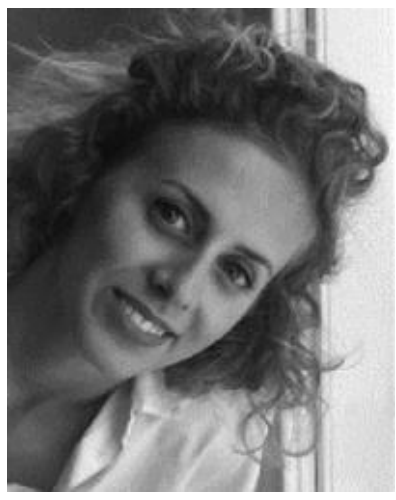

**MICHELA LONGHI** received the B.S. and M.S. degrees (summa cum laude) in electronic engineering from the Università Politecnica delle Marche, Ancona, Italy, in 2012 and 2014, respectively and the Ph.D. degree with European label in 2019, from the University of Tor Vergata, Rome, Italy. In 2014, she joined the European Space Agency, Noordwijk, The Netherlands, for the master thesis project. In 2015, she was Research Fellow at Consiglio Nazionale delle Ricerche (CNR-IEIIT), Milan, Italy, in the group of Engineering for Health and Weelbeing. In 2019, she was hosted by the ETH Zurich, Switzerland, for Ph.D. internship. She worked as RF and System Engineering at MVG Italy, Pomezia, Italy from 2019 to 2021. She is currently a Postdoctoral Researcher at the Niccolò Cusano University, Rome, Italy. Her research interests include beam array antennas, deep transcranial magnetic simulation, radiofrequency identification devices, drones, antennas designd, tests and measurements and recently, AI applications for metamaterials design.

Dr. Longhi is currently a member of the Italian Society on Electromagnetics (SIEM), of the National Inter-University Consortium for Telecommunications (CNIT) and of the Virtual Institute for Artificial Electromagnetic Materials and Metamaterials (Metamorphose VI AISBL).

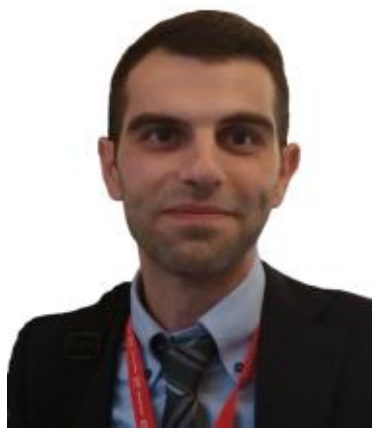

**ALESSIO MONTI** (S'12–M'15–SM'19) was born in Rome, Italy, in 1987. He received the B.S. degree in electronic engineering (summa cum laude), the M.S. degree in telecommunications engineering (summa cum laude) and the Ph.D degree in biomedical electronics, electromagnetics and telecommunications engineering from ROMA TRE University, Rome, Italy, in 2008, 2010 and 2015, respectively.

From 2013 to 2021, he was with the Niccolò Cusano University, Rome, Italy, while, since November 2021, he is with the ROMA TRE University where he serves as an Associate Professor in Electromagnetic Field Theory.

Prof. Monti is member of the secretarial office of the International Association METAMORPHOSE VI, and of the Editorial Board of the journals EPJ Applied Metamaterials (since 2016) and IEEE Transaction on Antennas and Propagation (since 2018). In 2019, he has been appointed as General Chair of the International Congress on Artificial Materials for Novel Wave Phenomena – Metamaterials and he has been serving as Chair of the Steering Committee of the same Congress series since 2017. He has been also member of the Technical Program Committee (TPC) of the IEEE International Symposium on Antennas and Propagation (2016-2019) and of the International Congress on Advanced Electromagnetic Materials in Microwaves and Optics-Metamaterials (2014-2016) and has been guest-editor of five journal special issues focused on metamaterials and nanophotonics. He has also been serving as a Technical Reviewer of many high-level international journals related to electromagnetic field theory, metamaterials and nanophotonics and he been selected as one of the Top Reviewers by the Editorial Board of the IEEE Transactions on Antennas & Propagation from 2014 to 2019.

His research interests include varied theoretical and application-oriented aspects of metamaterials and metasurfaces at microwave and optical frequencies, the design of functionalized covers and invisibility devices for antennas and antenna arrays and the electromagnetic modelling of micro- and nano-structured artificial surfaces. His research activities resulted in 100+ papers published in international journals, conference proceedings and book chapters. Prof. Monti has been the recipient of several national and international awards and recognitions, including the URSI Young Scientist Award (2019), the outstanding Associate Editor of the IEEE Transactions on Antennas and Propagation (2019&2020), the Finmeccanica Group Innovation Award for young people (2015) and the 2nd place at the student paper competition of the conference Metamaterials' (2012).

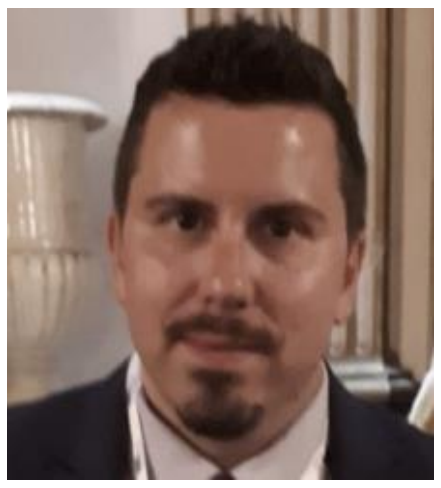

**DAVIDE RAMACCIA** (S'11–M'14–SM'19) received the B.S. (summa cum laude) and M.S. (summa cum laude) degrees in electronic and ICT engineering and the Ph.D. degree in electronic engineering from ROMA TRE University, Rome, Italy, in 2007, 2009, and 2013, respectively. Since 2013, he has been with the Department of Engineering (2013-2021) and, then, with the Department of Industrial, Electronic, and Mechanical Engineering (2021-now), at ROMA TRE University. His main research interests are in the the modelling and design of (space-)time-varying metamaterials and metasurfaces, and their applications to microwave components and antennas, and the analysis of anomalous scattering effects in temporal metamaterials. He has coauthored more than 100 articles in international journals, conference proceedings, book chapters, and holds one patent.

Davide Ramaccia has been serving the scientific community, by playing roles in the management of scientific societies, in the editorial board of international journals, and in the organization of conferences and courses. He is currently a General Secretary of the Virtual Institute for Artificial Electromagnetic Materials and Metamaterials (METAMORPHOSE VI, the International Metamaterials Society) and is an Elected Member of the Board of Directors of the same association from three consecutive terms (2014-now). Davide Ramaccia serves as an Associate Editor for the IEEE Access (2019-now), a Scientific Moderator for IEEE TechRxiv (2019-now), a Technical Reviewer of the major international journals related to electromagnetic field theory and metamaterials. He was also as Guest co-editor of three special issues on metamaterials and metasurfaces.

Since 2015, he serves as a member of the Steering Committee of the International Congress on Advanced Electromagnetic Materials in Microwaves and Optics – Metamaterials Congress. He has been General Chair and Local organizer of the 39$^{th}$ and 42$^{nd}$ EUPROMETA doctoral school on metamaterials held in Rome, Italy, in 2019 and 2021, respectively. He has been Technical Program Coordinator (Track "Electromagnetics and Materials") for the 2016 IEEE Antennas and Propagation Symposium. He is member of the Technical Program Committee of the International congress on Laser science and photonics applications - CLEO 2022. He has also been elected as a Secretary of the Project Management Board of the H2020 CSA project NANOARCHITECTRONICS (2017–2018).

Davide Ramaccia was the recipient of a number of awards and recognitions, including The Electromagnetics Academy Young Scientist Award (2019) seven Outstanding Reviewer Award by the IEEE Transactions on Antennas and Propagation (2013-2021), the IET prizes for the best poster on microwave metamaterials (2013) and IET Award for the Best Poster on the Metamaterial Application in Antenna Field (2011).

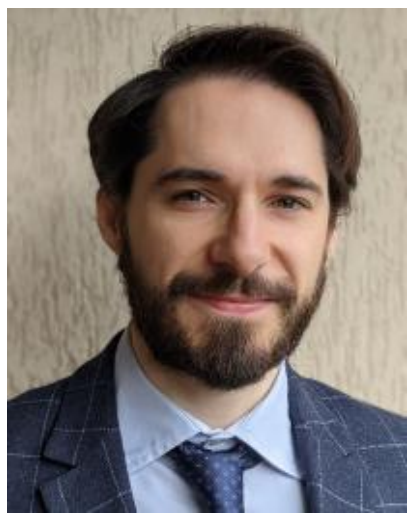

**STEFANO VELLUCCI** (S'16–M'20) received the B.S. and M.S. (summa cum laude) degrees in Electronic Engineering, and the Ph.D. degree in Applied Electronics from ROMA TRE University, Rome, Italy, in 2012, 2015, and 2019, respectively. He is currently a Postdoctoral Research Fellow with the Department of Industrial, Electronic, and Mechanical Engineering, at ROMA TRE University.

In 2014 he was an intern with MBDA, Missile Systems, Rome, Italy, and in 2015 he was an Antenna Engineer with ELT Elettronica S.p.A. Rome, Italy, where he designed, modeled, and optimized antennas for military applications. From 2019 to 2020 he was a Postdoctoral Researcher with the ELEDIA Research Center at the University of Trento, Trento, Italy, involved in the study and development of metasurfaces for space and terrestrial applications. His current research interests include the design and applications of artificially engineered materials and metamaterials to RF and microwave components, non-linear and reconfigurable circuit-loaded metasurfaces for radiating structures, analysis and design of metasurface-based cloaking devices for the antennas.

Dr. Vellucci is currently serving as Associate Editor for the EPJ Applied Metamaterials journal, has been guest-editor of two journal special issues focused on microwave, photonic, and mechanical metamaterials, and is member of the Virtual Institute for Artificial Electromagnetic Materials and Metamaterials (METAMORPHOSE VI), the Institute of Electrical and Electronics Engineers (IEEE), and the Italian Society on Electromagnetics (SIEM). In 2019 he was a local organizing committee member of the International Congress on Artificial Materials for Novel Wave Phenomena – Metamaterials, and since 2016 he has been serving as a Technical Reviewer of many high-level international journals and conferences related to electromagnetic field theory, metamaterial, and metasurfaces. He was the recipient of some national and international awards including the IEEE AP-S Award of the Central-Southern Italy Chapter (2019), the Outstanding Reviewers Award assigned by the IEEE Transactions on Antennas and Propagation (2018-2019-2020), and the Leonardo-Finmeccanica Innovation Award for "Young Students" (2015).

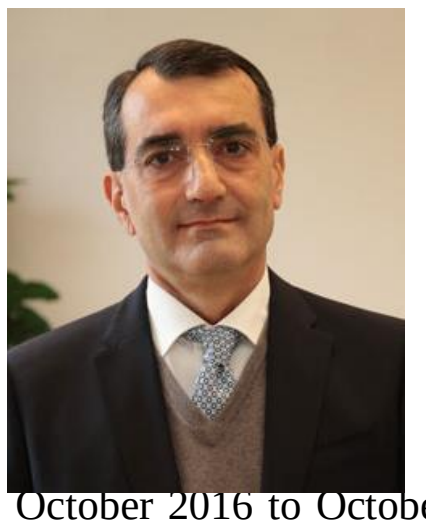

**ALESSANDRO TOSCANO** (M'91–SM'11) (Capua, 1964) graduated in Electronic Engineering from Sapienza University of Rome in 1988 and he received his PhD in 1993. Since 2011, he has been Full Professor of Electromagnetic Fields at the Engineering Department of Roma Tre University. He carries out an intense academic and scientific activity, both nationally and internationally.

From April 2013 to January 2018 he was a member of Roma Tre University Academic Senate. From October 2016 to October 2018, he is a member of the National Commission which enables National Scientific Qualifications to Full and Associate Professors in the tender sector 09/F1 – Electromagnetic fields.

Since 23rd January 2018 he has been Vice-Rector for Innovation and Technology Transfer.

Prof. Toscano is currently member of the board of director of Radiolabs (a non-for-profit Research Consortium), of the steering committee of the National Competence Center on Cyber 4.0, and of the scientific council of CIRIAF (Interuniversity Research Center on Pollution and the Environment).

In addition to his commitment in organizing scientific events, he also carries out an intense editorial activity as a member of the review committees of major international journals and conferences in the field of applied electromagnetics. He has held numerous invited lectures at universities, public and private research institutions, national and international companies on the subject of artificial electromagnetic materials, metamaterials and their applications. He actively participated in founding the international association on metamaterials Virtual Institute for Advanced Electromagnetic Materials – METAMORPHOSE, VI. He coordinates and participates in several research projects and contracts funded by national and international public and private research institutions and industries.

Alessandro Toscano's scientific research has as ultimate objective the conceiving, designing and manufacturing of innovative electromagnetic components with a high technological content that show enhanced performance compared to those obtained with traditional technologies and that respond to the need for environment and human health protection. His research activities are focused on three fields: metamaterials and unconventional materials, in collaboration with Professor A. Alù's group at The University of Texas at Austin, USA, research and development of electromagnetic cloaking devices and their applications (First place winner of the Leonardo Group Innovation Award for the research project entitled: 'Metamaterials and electromagnetic invisibility') and the research and manufacturing of innovative antenna systems and miniaturized components (first place winner of the Leonardo Group Innovation Award for the research project entitled: "Use of metamaterials for miniaturization of components" – MiniMETRIS).

He is the author of more than one hundred publications in international journals indexed ISI or Scopus; of these on a worldwide scale, three are in the first 0.1 percentile, five in the first 1 percentile and twenty-five in the first 5 percentile in terms of number of quotations and journal quality.

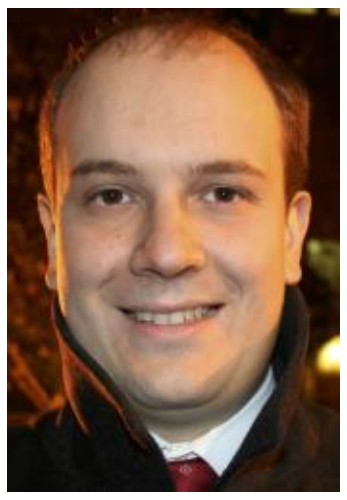

**FILIBERTO BILOTTI** (S'97–M'02–SM'06–F'17) received the Laurea and Ph.D. degrees in electronic engineering from ROMA TRE University, Rome, Italy, in 1998 and 2002, respectively. Since 2002, he has been with the Faculty of Engineering (2002-2012), the Department of Engineering (2013-2021), and the Department of Industrial, Electronic, and Mechanical Engineering (2021-now) at ROMA TRE University, where he serves as a Full Professor of electromagnetic field theory (2014-now) and the Director of the Antennas and Metamaterials Research Laboratory (2012-now).

His main research contributions are in the analysis and design of microwave antennas and arrays, analytical modelling of artificial electromagnetic materials, metamaterials, and metasurfaces, including their applications at both microwave and optical frequencies. In the last ten years, Filiberto Bilotti's main research interests have been focused on the analysis and design of cloaking metasurfaces for antenna systems, on the modelling and applications of (space and) time-varying metasurfaces, on the topological-based design of antennas supporting structured field, on the modelling, design, implementation, and application of reconfigurable metasurfaces, on the concept of meta-gratings and related applications in optics and at microwaves, on the modelling and applications of optical metasurfaces. The research activities developed in the last 20 years has resulted in more than 500 papers in international journals, conference proceedings, book chapters, and 3 patents.

Prof. Bilotti has been serving the scientific community, by playing leading roles in the management of scientific societies, in the editorial board of international journals, and in the organization of conferences and courses. In particular, he was a founding member of the Virtual Institute for Artificial Electromagnetic Materials and Metamaterials – METAMORPHOSE VI in 2007. He was elected as a member of the Board of Directors of the same society for two terms (2007-2013) and as the President for two terms (2013-2019). Currently, he serves the METAMORPHOSE VI as the Vice President and the Executive Director (2019-now).

Filiberto Bilotti served as an Associate Editor for the IEEE Transactions on Antennas and Propagation (2013-2017) and the journal Metamaterials (2007-2013) and as a member of the Editorial Board of the International Journal on RF and Microwave Computer-Aided Engineering (2009-2015), Nature Scientific Reports (2013-2016), and EPJ Applied Metamaterials (2013-now). He was also the Guest Editor of 5 special issues in international journals.

He hosted in 2007 the inaugural edition of the International Congress on Advanced Electromagnetic Materials in Microwaves and Optics – Metamaterials Congress, served as the Chair of the Steering Committee of the same conference for 8 editions (2008-2014, 2019), and was elected as the General Chair of the Metamaterials Congress for the period 2015-2018. Filiberto Bilotti was also the General Chair of the Second International Workshop on Metamaterials-by-Design Theory, Methods, and Applications to Communications and Sensing (2016) and has been serving as the chair or a member of the technical program, steering, and organizing committee of the main national and international conferences in the field of applied electromagnetics.

Prof. Bilotti was the recipient of a number of awards and recognitions, including the elevation to the IEEE Fellow grade for contributions to metamaterials for electromagnetic and antenna applications (2017), outstanding Associate Editor of the IEEE Transactions on Antennas and Propagation (2016), NATO SET Panel Excellence Award (2016), Finmeccanica Group Innovation Prize (2014), Finmeccanica Corporate Innovation Prize (2014), IET Best Poster Paper Award (Metamaterials 2013 and Metamaterials 2011), Raj Mittra Travel Grant Senior Researcher Award (2007).